\documentclass[aps,prb,twocolumn,10pt,longbibliography]{revtex4-1}
\usepackage[english]{babel}
\usepackage{amsmath}
\usepackage{amsfonts}
\usepackage{amssymb}
\usepackage{epsfig}
\usepackage{color}
\usepackage{graphicx}
\usepackage{colortbl}
\usepackage[hyperindex=true]{hyperref}
\hypersetup{linktocpage,colorlinks=true,citecolor=blue,linkcolor=blue}
\usepackage[inline]{enumitem}
\usepackage{braket}
\usepackage{comment}
\usepackage[normalem]{ulem}
\usepackage{pdfpages}
\usepackage{pgffor}

\makeatletter
\AtBeginDocument{\let\LS@rot\@undefined}

\definecolor{blue}  {rgb}{ 0,   0,   1 }

\newcommand{\Omrie}[1]{\textcolor{blue}{#1}}
\newcommand{\grad}{\boldsymbol{\nabla}}

\newcommand{\abs}[1]{\left|#1\right|}
\DeclareMathOperator{\tr}{tr}
\DeclareMathOperator{\Tr}{Tr}
\DeclareMathOperator{\sgn}{sign}
\DeclareMathOperator{\rank}{rank}
\DeclareMathOperator{\Index}{Index}

 \let\Omrie\relax             
 \renewcommand{\sout}[1]{}    

\makeatother

\begin{document}

\title{Vacancies in Graphene : Dirac Physics and Fractional  Vacuum Charges}

\author{Omrie Ovdat}
\author{Yaroslav Don}
\author{Eric Akkermans}
\email{eric@physics.technion.ac.il}

\affiliation{Department of Physics, Technion -- Israel Institute of Technology, Haifa 3200003, Israel}


\begin{abstract}

The study of vacancies in graphene is a topic of growing interest. A single vacancy induces a localized stable charge of order unity interacting with other charges of the conductor through an unscreened  Coulomb potential. It also breaks the symmetry between the two triangular graphene sublattices hence inducing zero energy states at the Dirac points. Here we show the fractional and pseudo-scalar nature of this vacancy charge. A continuous Dirac model is presented which relates zero modes to vacuum fractional charge and to a parity anomaly. This relation constitutes an Index theorem and is achieved by using particular chiral boundary conditions, which map the vacancy problem onto edge state physics. Vacancies in graphene thus allow to realize prominent features of $2+1$ quantum electrodynamics but without coupling to a gauge field.
This essential difference makes vacancy physics relatively easy to implement and an interesting playground for topological \Omrie{state} switching.

\end{abstract}

\maketitle



\section{Introduction and statement of results} 
\begin{figure}[ht]
   \includegraphics[width=0.49\textwidth]{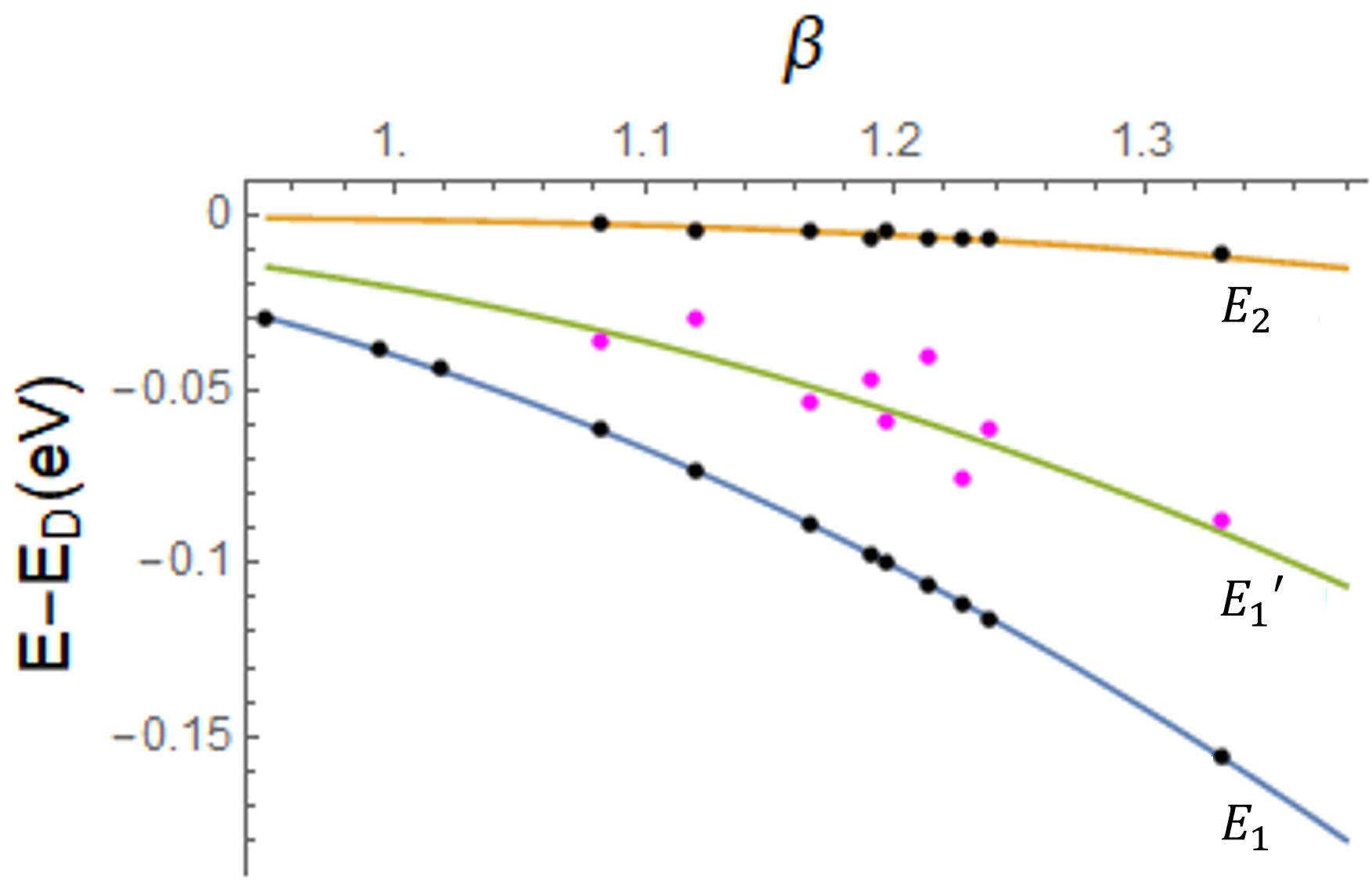}
\caption{\label{fig:CSI to DSI graphene} \textbf{Experimental observation of the massless Dirac-Coulomb spectrum in graphene with broken sublattice/parity symmetry (see  \cite{Ovdat2017} for more details).} \Omrie{The continuous lines above are dervied from the exact solution of the massless Dirac-Coulomb system where $\beta$ is the Coulomb strength and $E_D$ is the Dirac point. The curves  $E_1, E_{1}', E_2$ describe quasi bound states extracted from the total density of states of the $j = 1/2$ ($E_1,\,E_2$) and $j=-1/2$ ($E_1'$) total angular momentum channels. $E_1,\,E_2$ and $E_1'$ are also the lowest quasi bound states appearing for the corresponding $\beta > 1/2$ values in the plot. The black and magenta dots correspond to experimental points obtained at a charged vacancy in graphene. These are  obtained from tunneling conductance data measured as a function of tunneling voltage at the vacancy site. The existence of the middle branch is a clear signal for the degeneracy lifting between the two lowest angular momentum channels $j=\pm1/2$ and thereby an indication of parity symmetry breaking.}}
\end{figure}

Graphene has a remarkable low energy spectrum described by an effective Dirac model, whose interest resides in its ability to account for a wealth of fundamental aspects specific to massless Dirac fermions. Vacancies \cite{sutherland1986localization, Lieb1989, PerreiraDisorderInduced, PerreiraModelingPhysRevB.77.115109, nanda2012electronic, mao2016realization, liu2014determining, Ovdat2017,InuiPhysRevB.49.3190, PalaciosPhysRevB.77.195428, PadmanabhanPhysRevB.93.165403,   mao2017quantum, Ulybyshev2015PhysRevLett.114.246801,  Ugeda2010PhysRevLett.104.096804, Valencia2017PhysRevB.96.125431, Schindler2014PhysRevLett.113.186802, Peres2006PhysRevB.73.125411, Weik2016PhysRevB.94.064204}, obtained by removing neutral carbon atoms, have important consequences for the physics of graphene:
\begin{enumerate*}[label=(\roman*)]
\item Zero energy modes. In the presence of $N_{A}+N_{B}$ vacancies, where $N_{A}$ $\left(N_{B}\right)$ is the number of vacancies corresponding to sublattice $T_A$ $(T_B)$, the tight binding Hamiltonian has $\abs{N_{A}-N_{B}}$ zero energy eigenvalues with vanishing wave function on the minority sublattice \cite{sutherland1986localization, Lieb1989, PerreiraDisorderInduced, PerreiraModelingPhysRevB.77.115109, nanda2012electronic, mao2016realization}. 
\item Charge. Density functional theory calculations \cite{liu2014determining} show that when a carbon atom is removed, the induced electronic rearrangement leads to a lower energy configuration and to an overall local electric charge in the ground state. In addition, tunnelling and Landau level spectroscopy \cite{mao2016realization} provide experimental support for the existence of this local charge and show, with very good agreement, an energy spectrum corresponding to an unscreened $V \sim -1/r \,$ Coulomb potential \Omrie{(see Fig.~\ref{fig:CSI to DSI graphene})}. 
\item Symmetry breaking. For $N_{A}\neq N_{B}$, sublattice symmetry is broken and so is parity in the continuum limit. For a single vacancy, the degeneracy lifting between the two lowest angular momentum channels $j=\pm1/2$, a clear indication of parity symmetry breaking, has been indeed observed  \Omrie{(see Fig.~\ref{fig:CSI to DSI graphene})}. 
\end{enumerate*}


In this paper, we present a continuous Dirac model of graphene, valid at low energy and applicable to an  arbitrary configuration of isolated vacancies,  which accounts for the above features and shows their direct relation. 
The localized, fractional and pseudo-scalar nature of the vacancy charge is a consequence of the asymmetry  between positive and negative parts of the spectrum as expressed by the occurrence of zero energy modes. This fractional charge does  not display Friedel-like density  oscillations and essentially differs from the screening resulting from the insertion of external charge defects \cite{ShytovPhysRevLett.99.236801,shytov2007atomic,PereiraPhysRevLett.99.166802,FoglerPhysRevB.76.233402,BiswasPhysRevB.76.205122,KolomeiskyPhysRevB.88.165428,DiVincenzoPhysRevB.29.1685}.  The vacuum charge density and its corresponding charge are obtained by solving the scattering problem of massless Dirac fermions by one vacancy while imposing on their wave function a new type of `chiral' boundary conditions. This choice unveils the topological nature of the  charge and its relation to zero modes under the form of an Index theorem.  \Omrie{We emphasize how the phenomena of a charged vacancy presented here, realizes the physics of fermion number fractionalisation  \cite{Jackiw1976PhysRevD.13.3398,HeegerRevModPhys.60.781,jackiwSchriefferNuclearPhysicsB, Redlich1984PhysRevLett.52.18,Niemi1983PhysRevLett.51.2077,Niemi1985PhysRevD.32.471,Boyanovsky1985PhysRevD.31.3234,Blankenbecler1985PhysRevD.31.2089,Blankenbecler1986PhysRevD.34.612,Jackiw1984PhysRevD.29.2375,Jaroszewicz1986PhysRevD.34.3128,hagen1990aharonov,Semenoff1984PhysRevLett.53.2449,HaldanePhysRevLett.61.2015, Fradkin1987PhysRevLett.57.2967,Chamon2008PhysRevB.77.235431,Jackiw1983}  with the topological content of the magnetic flux $\Phi$ now replaced by vacancies with properly chosen boundary conditions such that
\begin{equation}
N_A - N_B  \leftrightarrow \Phi.
\end{equation}
}

\section{Dirac model} 
In graphene, carbon atoms condense into a planar honeycomb bipartite lattice built from two triangular sublattices $T_A$ and $T_B$. The Bravais lattice with a two-atom unit cell and its reciprocal are triangular  and the hexagonal Brillouin zone has two inequivalent crystallographic Dirac points $K$ and $K' $. Around each of them, the low energy excitation spectrum  is conveniently described by non-interacting and in-plane massless Dirac fermions with the effective continuous Hamiltonian,
\begin{equation}
H = - i \, \boldsymbol{\sigma} \cdot \boldsymbol{ \nabla} = 
\begin{pmatrix}
	0          & D  \\
	D^\dagger  & 0 
\end{pmatrix}
\label{eq: dirac}
\end{equation}
($\hbar = v_{\text{F}} = 1$),  $D = -i\partial_x -  \partial_y = e^{-i \theta} \left( - i \partial_r - \frac{1}{r} \partial_\theta \right)$ and $\boldsymbol{\sigma}=\left(\sigma_{x},\sigma_{y}\right)$. This description was shown to be valid at low energies even in the presence of electron-electron interactions up to logarithmic corrections to the Fermi velocity \cite{RevModPhys.84.1067, Elias2011} (see Supplementary Note 1).
The operators $D$ and $D^\dagger$ are defined on the direct sum $ {\mathcal H}_A \oplus {\mathcal H}_B$ of Hilbert spaces associated to $T_A$ and $T_B $ and the corresponding quantum states are two-component spinors $\psi\left(\boldsymbol{r}\right) = \begin{pmatrix}\psi^A &
\psi^B \end{pmatrix}^T$, with $\psi^{A,B}$ being quantum amplitudes on $T_A$ and $T_B$ respectively at a coarse grained position $\boldsymbol{r}$. 
%
%
The spectrum of $H$ spans the continuum, but positive and negative parts can be mapped one onto the other, a symmetry expressed by 
\begin{equation} 
\{ H,\sigma_z \} = 0 \, , 
\label{eq: sigma3} 
\end{equation}
and hereafter called chiral, which is a consequence of the bipartite structure of the lattice.
Moreover, the honeycomb lattice is invariant under spatial inversion $\boldsymbol{r} \mapsto - \boldsymbol{r}$
which decomposes into two mirror symmetries where parity, 
\begin{equation}
x\mapsto x,\,y\mapsto-y,  H\mapsto\sigma_{x}H\sigma_{x},
\end{equation}
 exchanges the two sublattices $T_A $ and $ T_B$.


The vacuum charge density, 
\begin{align}
\rho \left( \boldsymbol{r}\right) & = -e  \sum_{n,E_{n}<0}\psi_{n}^{\dagger}\left(\boldsymbol{r}\right)\psi_{n}\left(\boldsymbol{r}\right) \nonumber \\
& \quad + e \sum_{n,E_{n}<0}\psi_{n}^{\dagger}\left(\boldsymbol{r}\right)\psi_{n}\left(\boldsymbol{r}\right) \Big\vert_{\text{free}},
\end{align}
 \sout{corresponding} \Omrie{corresponds} to \Omrie{the particle density associated with} electrons filling all the negative energy states \Omrie{relative to the same quantity in absence of any potentials. Utilizing the completeness relation $\rho \left( \boldsymbol{r}\right)$} 
takes the symmetric form \cite{Stone1985PhysRevB.31.6112,Niemi1984PhysRevD.30.809},
\begin{equation}
\rho \left( \boldsymbol{r}\right) = \frac{e}{2} \sum_n \sgn \left(E_n\right) \psi_{n}^{\dagger}\left(\boldsymbol{r}\right)\psi_{n}\left(\boldsymbol{r}\right) .
\label{eq: eta}
\end{equation}
For an infinite system, the charge density $ \rho \left( \boldsymbol{r}\right)$ is a total divergence (see \cite{Niemi1984PhysRevD.30.809,callias1978axial} and Supplementary Note 2),
\begin{equation}
\rho \left( \boldsymbol{r}\right) = \frac{e}{2} \sgn \left(M\right) \boldsymbol{\nabla} \cdot  \boldsymbol{\Delta} (\boldsymbol{r}) 
\label{eq: eta2}
\end{equation}
where the regularising mass parameter $M\to 0$ removes the sign ambiguity in \eqref{eq: eta} in the presence of zero modes. \Omrie{The ambiguity associated with $E = 0$ results from the necessity to determine whether
or not $E = E_F = 0$ states are occupied. The introduction of a small mass
term is one way to regularize this ambiguity. The mass term shifts the zero modes
to $\pm M$ which, depending on the sign, discriminates between occupying
the zero modes or not.} The matrix element 
\begin{equation}
\boldsymbol{\Delta} (\boldsymbol{r }) \equiv \frac{1}{2}  \bigl\langle \boldsymbol{r} \big| \tr \bigl( \boldsymbol{\sigma} \sigma_z \frac{1}{H - i0} \bigr) \big| \boldsymbol{r} \bigr\rangle    
\end{equation}
 is a two-dimensional vector and ``tr''  is over spinor indices.

Despite being defined over the entire energy spectrum, $ \rho \left( \boldsymbol{r}\right)$ turns out to be related to a quantity evaluated at the Fermi energy, a noteworthy result since \eqref{eq: dirac} is merely valid close to $E=0$. Furthermore, \eqref{eq: eta2} is directly related to features of the zero-energy subspace. Its dimension, $\dim \ker D + \dim \ker D^\dagger$, obtained by counting all solutions of $D \psi_B = D^\dagger \psi_A =0$,  cannot  generally be determined, but the relation, 
\begin{equation}
 \Index H =  - \sgn \left( M \right) \int d \boldsymbol{r} \, \boldsymbol{\nabla} \cdot  \boldsymbol{\Delta} (\boldsymbol{r})
\label{eq: index}
\end{equation}
holds for $\Index  H  \equiv \dim \ker D - \dim \ker D^\dagger$ \cite{Niemi1984PhysRevD.30.809,Stone1985PhysRevB.31.6112}. Combining \eqref{eq: eta2} and \eqref{eq: index} leads to  
\begin{equation}
Q \equiv \int d \boldsymbol{r} \rho \left( \boldsymbol{r}\right) = - \frac{e}{2}  \Index H \, .
\label{eq: pseudocharge}
\end{equation}
In the absence of vacancies, there are no zero modes thus $\Index H $ vanishes and so does the charge $Q$ and $\rho \left( \boldsymbol{r} \right)$. However, this may not be the case in the presence of vacancies.


\section{Scattering description of single vacancy}
The removal of one carbon atom creates a vacancy, here arbitrarily assigned to be an $A$-vacancy
\footnote{In the presence of a defect, the Dirac points $K$ and $K'$ may be coupled, which is inconsistent with the single valley continuum Dirac equation. As shown in \cite{Ando1998},
for the case a Gaussian scatterer with range $R=a$, the ratio between
the inter-valley and intra-valley coupling is small 
(around $\sim0.05$). Thus, we neglect this coupling here, an assumption that will be further justified numerically}. 
The corresponding excitation spectrum in the continuum limit is obtained by considering scattering solutions of the Dirac Hamiltonian \eqref{eq: dirac} on a plane with a puncture of radius $R$.  \sout{Generically, the spectral imprint of a small defect is expected to be significant only near zero energy, which is consistent with the Dirac picture}. Since $\rho \left( \boldsymbol{r}\right)$  depends on the behaviour at zero energy, we look for zero modes,  i.e., solutions of $D \psi_B = D^\dagger \psi_A =0$. The general solution is
\begin{equation}
\psi \left(r,\theta\right) \equiv  \sum_{m \in \mathbb{Z}} e^{i m \theta}  
\begin{pmatrix}
	  \psi_m ^A (r) \quad   \\
	i \psi_m ^B (r) e^{i \theta}
\end{pmatrix}
\label{eq: zeromode1}
\end{equation}
with $\psi_m ^A (r) =  A_m r^m  $,  $\psi_m ^B (r) = 
 B_m r^{-m-1} $ and $\left(A_m , B_m \right)$ constants. Requiring $\psi \left(r \rightarrow  \infty, \theta \right) = 0$, we keep harmonics $m <0$ for $\psi_m ^A (r)$ and $m \geq 0$ for $\psi_m ^B (r)$. 

\subsection{Chiral boundary conditions}
We choose  appropriate boundary conditions on the scattering potential. \sout{ so as to preserve chiral symmetry \eqref{eq: sigma3}, a necessary condition to use expressions \eqref{eq: eta2}--\eqref{eq: pseudocharge}}. Local boundary conditions e.g., Dirichlet, $\psi \left(\boldsymbol{r}\right) |_\text{vac} =0$ lead either to an over determination or to particle-hole pair creation (Neumann) \cite{BerryRoyalSociety412.1842}. We propose instead a new set of  chiral boundary conditions, 
\begin{equation}
\begin{aligned}
	\psi_m ^A (r=R) & = 0, & m & \leq0,  \\
	\psi_m ^B (r=R) & = 0, & m & > 0 ,
\end{aligned}
\label{eq: chiralbcs}
\end{equation}
a close relative of non-local boundary conditions introduced in the study of Index theorems for Dirac operators \cite{atiyah1975spectral,AkkermansPhilosophicalMagazineB77.5,AkkermansEurPhyB1.1}. This choice (\ref{eq: chiralbcs}) preserves the chiral symmetry \Omrie{\eqref{eq: sigma3}, a necessary condition to use expressions \eqref{eq: eta2}--\eqref{eq: pseudocharge},}
and represent a perfectly reflecting barrier of probability density (Supplementary Note 3). Implemented on the  power law wave function \eqref{eq: zeromode1}, conditions \eqref{eq: chiralbcs} uniquely lead to a single zero mode 
\begin{equation}
\psi\left(\boldsymbol{r}\right) \equiv \begin{pmatrix}0 \\ i B_0  e^{i \theta}/ r \end{pmatrix}    
\end{equation}
 by projecting onto the $m=0$ subspace for $\psi_m ^B (r)$ and having $\psi_m ^A \equiv 0$.  
It is worth noting that this eigenfunction reproduces the tight binding result \cite{PerreiraDisorderInduced} justified by the  absence of any characteristic scale. This zero mode is quasi-bound, that is, decaying but non-normalizable and thus appears as a pronounced peak in the density of states at the Fermi energy. An analogous choice of boundary conditions for a $B$-vacancy \Omrie{, presented in Tab. \ref{tab:BC table},} leads  to the single zero mode $\psi\left(\boldsymbol{r}\right) \equiv \begin{pmatrix}  A_{-1}/r  & 0 \end{pmatrix}^T$ \footnote{There exists alternative choices of chiral boundary conditions e.g., by projecting the zero modes onto a different angular momentum subspace. However, this would yield a zero mode decaying faster than $1/r$ and with a dimension-full strength\label{footnote: different AM}}.

\begin{table}[t]
\caption{\label{tab:BC table} \textbf{Boundary condition.} Boundary conditions for an $A/B$-vacancy imposed
on the radial components $\psi_m^{A,B}$.
The conditions differ only for $m = 0,-1$ ($j = \pm1/2$).}
\smallskip
\begin{tabular}{rcccccc}
\hline 
 && \multicolumn{2}{c}{$A$-vacancy} && \multicolumn{2}{c}{$B$-vacancy}\tabularnewline
\noalign{\vskip\doublerulesep}
$m$ \:{} && $\psi_{m}^{A}\left(R\right)$ & $\psi_{m}^{B}\left(R\right)$ && $\psi_{m}^{A}\left(R\right)$ & $\psi_{m}^{B}\left(R\right)$\tabularnewline
\hline 
\noalign{\vskip\doublerulesep}
$\leq-2$ \;{} && $0$ &  && $0$ & \tabularnewline
\rowcolor{yellow}$-1$ \;{} && $0$ &  &&  & $0$\tabularnewline
\rowcolor{yellow}$0$ \;{} && $0$ &  &&  & $0$\tabularnewline
$\geq1$ \;{} &&  & $0$ &  && $0$\tabularnewline
\hline 
\end{tabular}
\end{table}

\subsection{Parity symmetry breaking}
\Omrie{As required by sublattice symmetry breaking}, chiral boundary conditions \eqref{eq: chiralbcs} do not preserve parity
which in the continuous limit, corresponds to $m \leftrightarrow -m -1$, $\psi_m ^A \leftrightarrow -\psi_{-m-1} ^B$ and $\psi_m ^B \leftrightarrow  \psi_{-m-1} ^A$. Indeed, unlike the parity preserving choice,
\begin{equation}
\begin{aligned}
\psi_m ^A (r=R) & = 0, & m & >0 , \\
\psi_m ^B (r=R) & = 0, & m & \leq 0 ,
\end{aligned}
\label{eq: BCond}
\end{equation}
under conditions \eqref{eq: chiralbcs}, the $m=0$ solution $\psi_{0} ^{B} (r) = i  e^{i \theta}/ r$ does not transform into the  vanishing $m =-1$ solution $\psi_{-1} ^A (r)$. We thus conclude that  the presence of a vacancy necessarily breaks parity and removes the $j =\pm 1/2$ degeneracy, where $j \equiv m + 1/2$. \Omrie{This point is particularly relevant in light of recent observation of $j = \pm 1/2$ degeneracy breaking by STM spectroscopy at a vacancy site \cite{Ovdat2017} (Fig. \ref{fig:CSI to DSI graphene})}.

\subsection{Results - single vacancy}
\begin{figure}
\centering
\includegraphics[width=0.98\columnwidth]{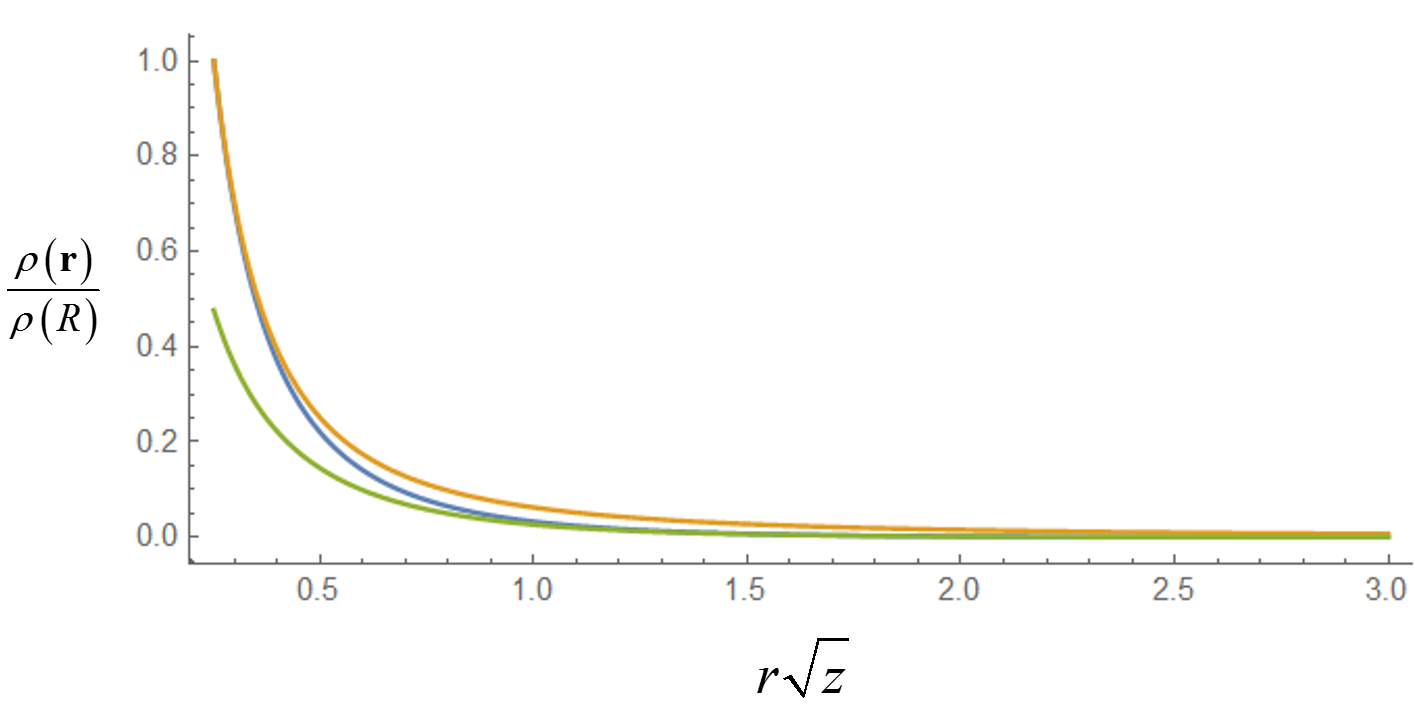}
\caption{\label{fig: chargeDesSingleVac} \textbf{Single vacancy charge density.} Blue: Characteristic behaviour of $\rho \left( \boldsymbol{r} \right)/\rho \left( R \right)$ in \eqref{eq: rho bessels} as a function of $x \equiv r \sqrt{z}$ with $y \equiv R \sqrt{z} = 0.25$. Orange: the function $y^2/x^2$. Green: The function $\pi y^2 \, e^{-2 x}/x$ }
\end{figure}

To relate the existence of the zero mode to a finite vacuum charge density as given in \eqref{eq: index}--\eqref{eq: pseudocharge}, we must directly calculate the Index in \eqref{eq: index}. To that aim, we use the  regularized expression \cite{callias1978axial},
\begin{equation}
\Index H = \lim_{z \rightarrow 0}  \,  \Tr  \left( \frac{z}{H^B + z} - \frac{z}{H^A + z} \right) 
\label{eq: callias}
\end{equation}
where $H^B \equiv D^\dagger D$ and $H^A \equiv D D^\dagger $. The  ``$\Tr$'' operation here is over all states. Hereafter we take $\sgn M \equiv 1$ in \eqref{eq: index}, thus arbitrarily fixing the sign of the charge for an $A$-vacancy. Extending chiral boundary conditions \eqref{eq: chiralbcs} to non-zero energy scattering states involved in \eqref{eq: callias}, shows how the angular momentum contributions cancel out except for $j = \pm 1/2 \leftrightarrow m=-1,0$.
A thorough calculation (Supplementary Note 4) yields

\begin{equation}
\Index H= - \frac{1}{2\pi R}\lim_{z\rightarrow0}\int d \boldsymbol{r} \, \boldsymbol{\grad}\cdot\left(\frac{K_{0}(\sqrt{z}r)K_{1}(\sqrt{z}r)}{K_{0}(\sqrt{z}R)K_{1}(\sqrt{z}R)}\hat{r}\right)\label{eq:Index equation}
\end{equation}
 where $K_{n}\left(x\right)$ are the modified Bessel functions of the second kind. Integrating \eqref{eq:Index equation} in the region $R<r<\infty,\,0<\theta<2\pi$
and inserting into \eqref{eq: pseudocharge} gives
\begin{equation}
Q=-\frac{e}{2}\Index H=-\frac{e}{2}\cdot\left(\lim_{z\rightarrow0}1\right)=-\frac{e}{2} \cdot 1.\label{eq:index =00003D 1}
\end{equation}
The charge density $\rho\left(\boldsymbol{r}\right)$ can be read off the integrand \footnote{Note that it is not allowed to move the $\lim_{z\rightarrow0}$ through
the integral in \eqref{eq:Index equation}.} in \eqref{eq:Index equation} 
\begin{equation}
\rho\left(\boldsymbol{r}\right) = - \frac{e}{4\pi R} \boldsymbol{\grad}\cdot\left(\frac{K_{0}(\sqrt{z}r)K_{1}(\sqrt{z}r)}{K_{0}(\sqrt{z}R)K_{1}(\sqrt{z}R)}\hat{r}\right).\label{eq: rho bessels}
\end{equation}
In the limit of a pointlike vacancy, $R \to 0$, $\rho\left(\boldsymbol{r}\right)$ vanishes $\forall r \neq 0$. Since $\int d \boldsymbol{r}\, \rho\left(\boldsymbol{r}\right) = -e/2$, independent of $R$, $\rho\left(\boldsymbol{r}\right)$ can be represented by the $\delta$-function distribution
\begin{equation}
\lim_{R \to 0} \rho\left(\boldsymbol{r}\right) = - \frac{1}{2\pi} \boldsymbol{\grad}\cdot\left(\frac{e/2}{r}\hat{r}\right) \label{eq: point charge density}.
\end{equation}
For finite $R$, $\rho\left(\boldsymbol{r}\right)$ can be approximated from \eqref{eq: rho bessels} with an arbitrarily small finite value of $z$ acting as an IR cutoff. For $r \sqrt{z} \gg 1$, $\rho\left(r\right)/\rho\left(R\right)\approx \exp \left(-2\sqrt{z}r \right)$
and for $r \sqrt{z} \ll 1$, $R \sqrt{z} \ll 1$,  $\rho\left(r\right)/\rho\left(R\right)\approx R^{2}/r^{2}$. Thus, the charge density decays close to the vacancy as $\sim1/r^{2}$ and decays exponentially, far from the vacancy (see Fig.~\ref{fig: chargeDesSingleVac}).

\begin{figure}
   \includegraphics[width=0.46\columnwidth]{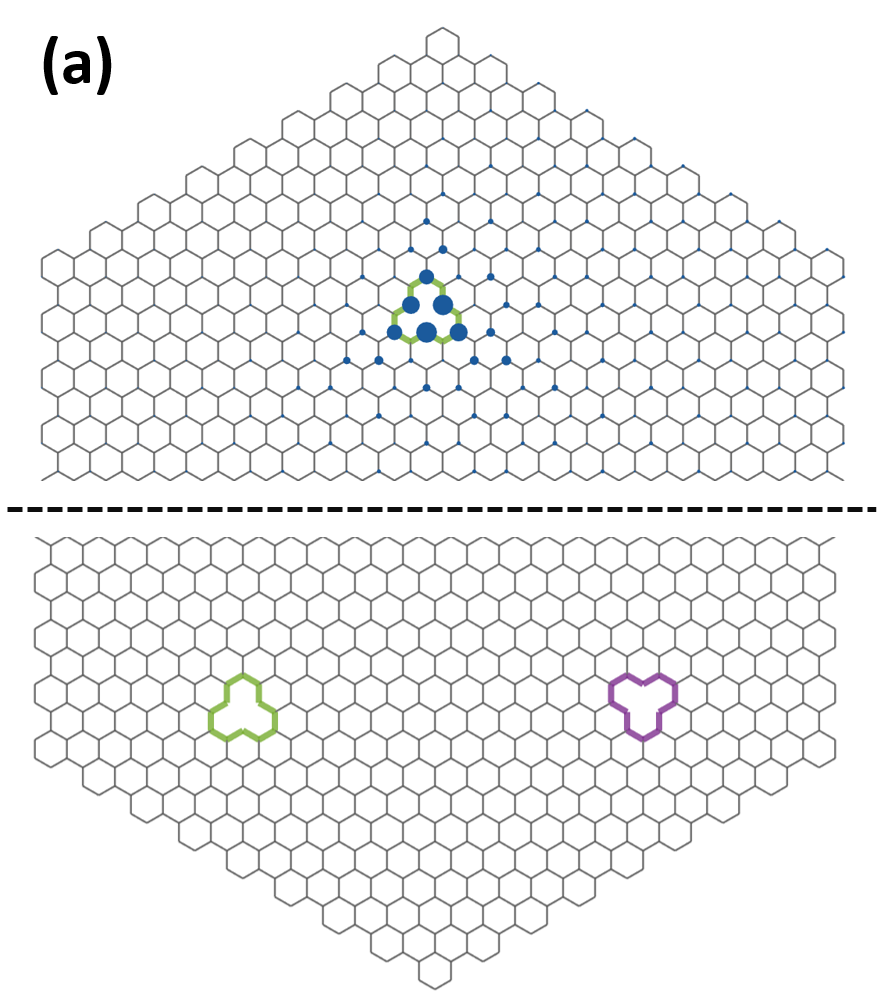}
   \includegraphics[width=0.52\columnwidth]{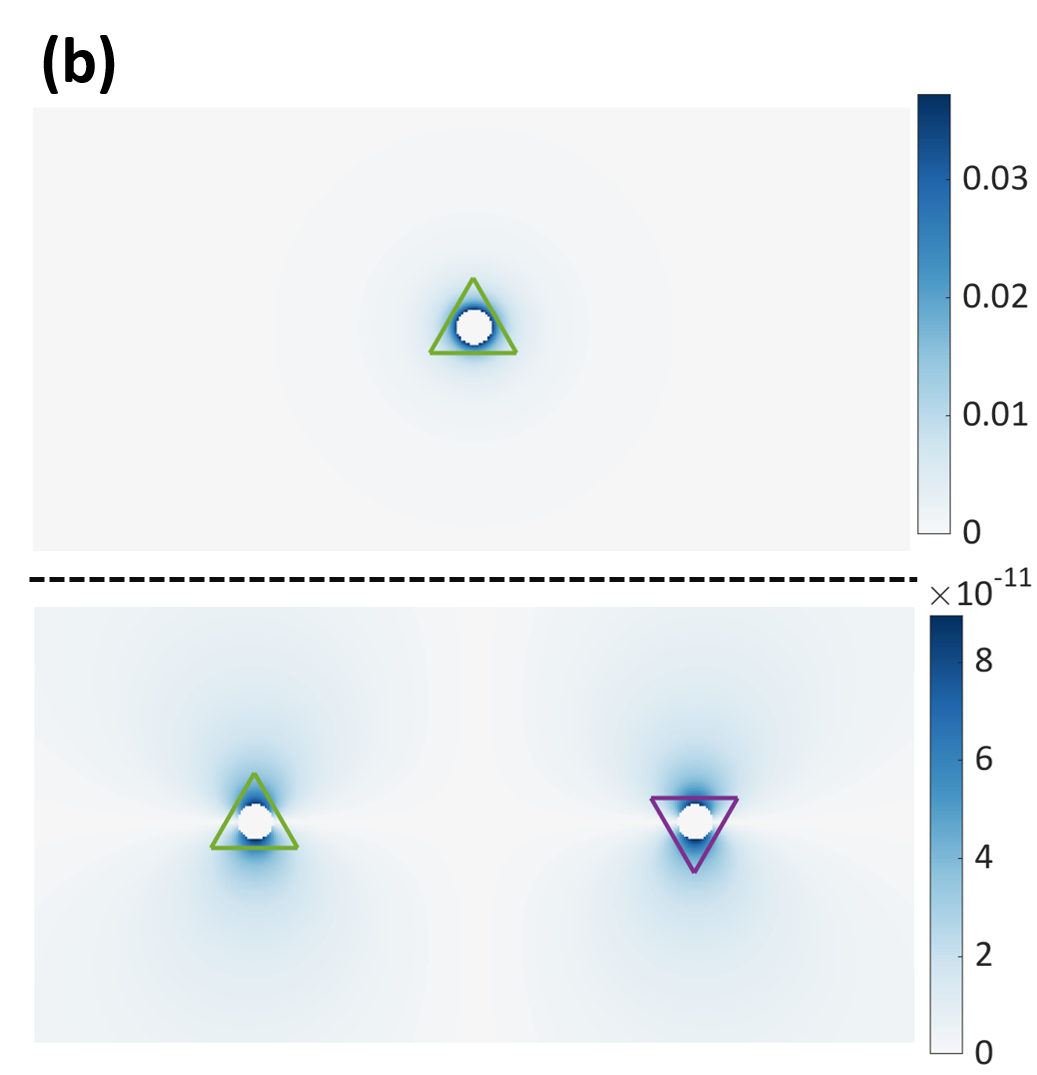}
   \caption{\label{fig: single and double} \textbf{Charge of vacancy configurations.} \textbf{Top:} Single $A$-vacancy ($N_A =1$, $N_B =0$). There is one zero mode, $\Index H = | N_A  - N_B  | =1$ and a finite fractional charge $Q = e/2$. \textbf{Bottom:}  $N_A = N_B =1$. Adding a $B$-vacancy, the zero mode disappears, $\Index H = | N_A  - N_B  | =0$, and so does the fractional charge on both vacancy locations represented for visual aid, by the green ($A$-vacancy) and purple ($B$-vacancy) outlines.
   \textbf{(a)} Tight binding calculation of the spatial charge density $\left|\rho \left( \boldsymbol{r} \right)\right|$ obtained from definition Eq.~(\ref{eq: eta}) and depicted by the blue spots. The total charge in a two lattice spacing radius is $Q \approx 10^{-1}$ (in units of ${e / 2}$) for the single vacancy (Top) and $Q_\vartriangle,\, Q_\triangledown \approx 10^{-8}$ for $N_A = N_B =1$ (Bottom). A small positive mass term $M \approx +10^{-9}$ has been used together with armchair boundary conditions which suppress charge accumulation on the boundary (Supplementary Note 5).
   \textbf{(b)} Continuous Dirac model calculation of the spatial charge density $\left|\rho \left( \boldsymbol{r} \right)\right|$ for the same situations as in \textbf{(a)}. These results are obtained using low energy scattering theory (Supplementary Note 6). Note the different scales displayed on the right color code.}
\end{figure}

\sout{Thus,} \Omrie{The resulting  charge density  $\rho\left(\boldsymbol{r}\right)$, is thus} \sout{ $\rho\left(\boldsymbol{r}\right)$ appears as} a total divergence with a fractional vacuum charge $ Q = - e / 2 $, localized at the  boundary of the vacancy (Fig.~\ref{fig: chargeDesSingleVac}, \ref{fig: single and double}). \Omrie{In the simplest approximation the corresponding potential, induced by electron interaction,} is Coulomb-like, i.e., decays as $1/r$. The same conclusions apply to a $B$-vacancy but with an opposite sign of the charge (Supplementary Note 4). This sign flip $Q \to -Q$ in the exchange $T_A \leftrightarrow T_B$ points to the pseudo-scalar nature of the vacuum charge. Hence a non-zero $Q$ provides a  clear signal for the breaking of parity symmetry of the ground state and  the lifting of the $j = \pm 1/2$ degeneracy. Including spin degeneracy, the overall ``fractional charge'' is $2 \times Q = \pm  e$.

It is interesting to further understand the origin of this finite charge.
The creation of a vacancy leads to an asymmetry between positive and negative energy states. An ill-defined albeit suggestive way to visualize it  is offered by the spatial integral of \eqref{eq: eta} which together with \eqref{eq: pseudocharge} gives  
\begin{equation}
Q = \frac{e}{2} \left(\sum_{E_{n}>0}1-\sum_{E_{n}<0}1\right) = - \frac{e}{2} \Index   H \, .
\label{eq: total charge}
\end{equation}
This ``spectral asymmetry'', of topological origin \cite{atiyah1975spectral}, eventually amounts to a counting of zero modes only.


All together, the fractional pseudo-scalar charge, the resulting Coulomb-like potential \footnote{The sign of the Coulomb potential depends on $\sgn M$ and whether the vacancy is from sublattice $T_A$ or $T_B$. The resulting spectrum 
remains unchanged in either case but will represent particle or hole states accordingly.} and the lifting of the $j = \pm 1/2$ degeneracy provide a comprehensive explanation to the observation of a vacancy charge and parity breaking obtained by STM measurements at a vacancy location in graphene  \cite{Ovdat2017}. Note that the charge density \eqref{eq: point charge density} does not display otherwise expected Friedel-like oscillations for the screening of a scalar charge.
%
%
\sout{This is yet another indication of the topological nature of the charge expressed by a finite Index \eqref{eq: pseudocharge}, resulting only from the choice of boundary conditions \eqref{eq: chiralbcs} and insensitive to both perturbations and specific  features of large energy dispersion law.} 
These findings thus constitute an original example of a non-zero Index in an open space,  independent of the existence of an underlying gauge field (above one spatial dimension). \sout{A finite $Q$ thus describes the physics of a topological defect.}



 
\section{Multiple vacancies} 
We now generalize the previous results to arbitrary configurations of a finite number of isolated vacancies. \Omrie{As in the single vacancy case, this description assumes non interacting electrons, corresponding to the Dirac and tight binding model of graphene. We discuss the validity of the associated multi-vacancy features in the discussion section.}

The zero mode wave functions are now difficult to obtain primarily due to multiple scattering between vacancies and the lack of rotational symmetry. Since the size of each vacancy is the lattice spacing, we assume constant wave function  
along the boundary of each vacancy making them point-like scatterers \cite{demkov2013zero}. Starting from the zero mode eigenfunctions,
\begin{equation}
\psi_\blacktriangle \left({z}\right) =\frac{1}{z^{\ast}-z_{A}^{\ast}}
\begin{pmatrix}0\\ 1\end{pmatrix},\quad
\psi_\blacktriangledown \left({z}\right)= \frac{1}{z-z_{B}}
\begin{pmatrix}1\\ 0 \end{pmatrix}
\label{eq: single vacancy solution}
\end{equation}
established for a single $A$ or $B$-vacancy located in $z_{A,B}$, $z \equiv x+iy$, we propose the ansatz,
\begin{equation}
\psi_{N}\left(z\right)=\begin{pmatrix}0\\ 1 \end{pmatrix}
\sum_{k=1}^{N_{A}}\frac{q_{kA}}{z^{\ast}-z_{kA}^{\ast}}
+\begin{pmatrix}1\\ 0 \end{pmatrix}
\sum_{k=1}^{N_{B}}\frac{q_{kB}}{z-z_{kB}}
\label{eq: many vacancy solution}
\end{equation}
for a configuration of $N = N_A + N_B$ vacancies 
located in  $z_{kA}$ and $z_{kB}$. This spinor wavefunction $\psi_N \equiv \begin{pmatrix}\psi_N ^A &
\psi_N ^B \end{pmatrix}^T$ reproduces all the single vacancy features previously obtained by means of chiral boundary conditions (\ref{eq: chiralbcs}), provided we require  $\psi_{N} ^A\left({z}_{kA}\right)= \psi_{N} ^B \left({z}_{kB}\right)=0$. The resulting constraints on the parameters $q_{kA,kB}$ take the matrix form,
\begin{equation}
\begin{aligned}
  M \boldsymbol{q}_{B}  & = 0,  & 
  M^{\dagger}\boldsymbol{q}_{A} & = 0,
\end{aligned}
\label{eq: q_A,B}
\end{equation}
where $M_{ij} = \left(z_{iA}-z_{jB}\right)^{-1}$ is a $N_{A}\times N_{B}$ Cauchy matrix of full rank $\forall z_{iA}, z_{jB}$ \cite{davis1975interpolation}.
Assuming, without loss of generality, that $N_{A}\geq N_{B}$, then $\rank M =\rank M^{\dagger} =N_{B}$ and
the solution of $M \boldsymbol{q}_{B}=0$ becomes the trivial one $\boldsymbol{q}_{B}=0$, while $M^{\dagger}\boldsymbol{q}_{A}=0$ has $N_{A}-N_{B}$ independent solutions, i.e., $\abs{N_{A}-N_{B}}$ zero modes for arbitrary $N_{A},N_{B}$. As \sout{expected} \Omrie{required}, this result coincides with the number of zero modes proven to exist in any vacancy filled bipartite lattice \cite{sutherland1986localization,Lieb1989,PerreiraDisorderInduced,PerreiraModelingPhysRevB.77.115109,nanda2012electronic}. Moreover, for $N_{A}\geq N_{B}$, all the zero modes fulfill $\psi_N^A \equiv 0$ and  $D\psi_N^B = 0$, thus, for a multi-vacancy configuration, $\Index = \# \text{ of zero modes} = N_A - N_B$. \Omrie{Utilizing scattering theory, we additionally obtained a closed form expression for $\rho \left( \boldsymbol{r}\right)$ as given in \ref{eq: eta2} for a general multi-vacancy configuration  (see Supplementary Note 6).} 


\begin{figure}
   \raisebox{-0.5\height}{\includegraphics[width=0.45\columnwidth]{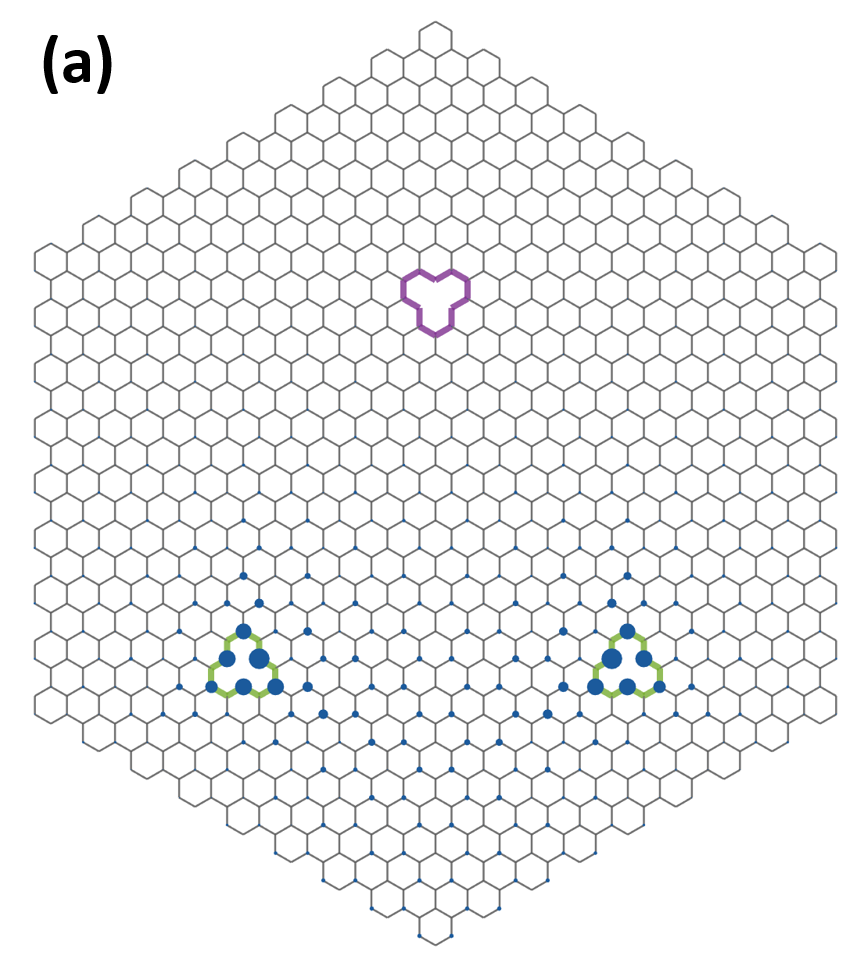}}
   \raisebox{-0.5\height}{\includegraphics[width=0.53\columnwidth]{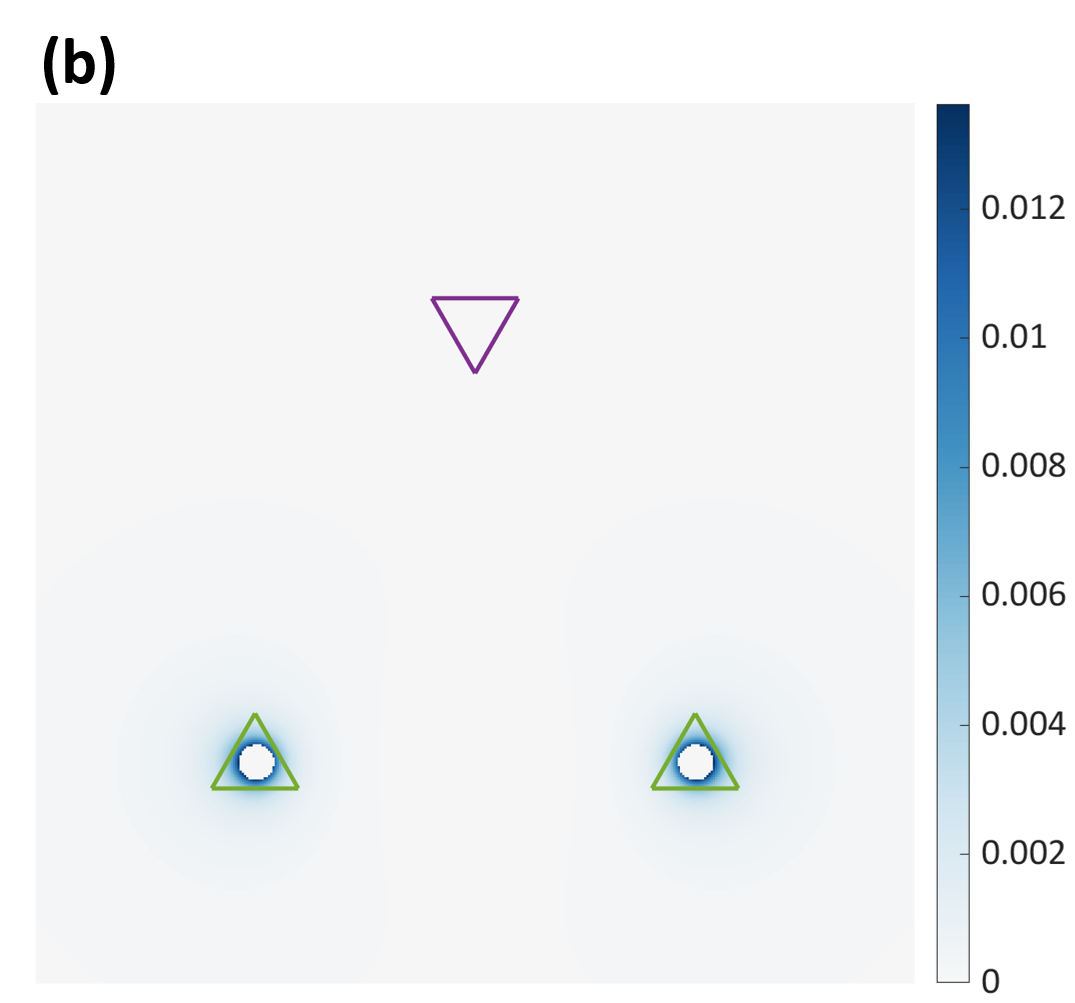}}
   \caption{\label{fig: TripleHoleCont} \textbf{Configuration of three vacancies $\boldsymbol{N_A =2}$, $\boldsymbol{N_B =1}$.} There is one zero mode, $\Index H = | N_A  - N_B  | =1$, so that the two $A$-vacancies (green upward outline) have a finite and equal charge $Q_\blacktriangle$ in this symmetric configuration and the $B$-vacancy (purple downward outline)  is not charged $Q_\triangledown =0$. 
   \textbf{(a)} Tight binding calculation of the spatial charge density $\left|\rho \left( \boldsymbol{r} \right)\right|$ obtained from definition Eq.~(\ref{eq: eta}) and depicted by the blue spots. The total charge is $Q_\blacktriangle \approx 10^{-1}$ (in units of ${e / 2}$) and $Q_\triangledown \approx 10^{-4}$ on each $A$,$B$ vacancy respectively. A small positive mass term $M \approx +10^{-9}$ has been used together with armchair boundary conditions which suppress charge accumulation on the boundary (Supplementary Note 5). 
   \textbf{(b)} Continuous Dirac model calculation of the spatial charge density $\left|\rho \left( \boldsymbol{r} \right)\right|$ for the same situation as in \textbf{(a)}. These results are obtained using low energy scattering theory (Supplementary Note 6). The homogeneous purple region around the $A$-vacancies is $\approx 10^{-5}$.}
\end{figure}

We now dwell on cases which illustrate the underlying features of many-vacancy configurations. \Omrie{In these cases we illustrate the correspondence of our Dirac model with tight binding numerics.} Starting from a single $A$-vacancy $(N_A =1)$ (Fig.~\ref{fig: single and double}). A zero mode appears associated to $\Index H = N_A =1$, together with a vacuum charge $Q = - (1/2) e$ localized at the vacancy site and a broken parity symmetry. Adding a $B$-vacancy (Fig.~\ref{fig: single and double}) implies $\Index H = | N_A  - N_B  |=0$ so that the charge vanishes at each vacancy location and parity symmetry is restored. 

Adding yet another $A$-vacancy changes the situation since  $\Index H = | N_A  - N_B  |=1$ and parity symmetry is again broken. Each $A$-vacancy now holds a finite  charge $Q_\blacktriangle $ smaller than $ (1/2) e$ which depends on the exact spatial configuration. The $B$-vacancy carries  no charge, $Q_\triangledown = 0$, a direct consequence of the vanishing of $\boldsymbol{q}_B$ in \eqref{eq: q_A,B}. 
These results, displayed in Fig.~\ref{fig: TripleHoleCont}, have a attractive generalisation. Consider a $N_A - N_B =1$ configuration where all the $A$-vacancies are charged ($Q_\blacktriangle$) and the $B$-vacancies necessarily uncharged ($Q_\triangledown$). Adding a $B$-vacancy wherever in the plane markedly changes this picture by switching off all the charges in the plane  ($Q_\vartriangle,  Q_\triangledown$). This feature, \sout{ illustrated in Fig.~\ref{fig: A5_B4 Vs A5_B5},} can be viewed as a topological state switch, where the creation of one  remote vacancy of the right kind switches off, at once, all the finite charges $Q_\blacktriangle$ on the graphene lattice. This effect is independent of the relative position of the vacancies and results only from the vanishing of the overall Index.




\section{Discussion}
The physics of a charged vacancy presented here, bears essential similarities with $2+1$ quantum electrodynamics (QED), such as fermion number fractionalisation and parity anomaly \cite{Jackiw1976PhysRevD.13.3398,HeegerRevModPhys.60.781,jackiwSchriefferNuclearPhysicsB, Redlich1984PhysRevLett.52.18,Niemi1983PhysRevLett.51.2077,Niemi1985PhysRevD.32.471,Boyanovsky1985PhysRevD.31.3234,Blankenbecler1985PhysRevD.31.2089,Blankenbecler1986PhysRevD.34.612,Jackiw1984PhysRevD.29.2375,Jaroszewicz1986PhysRevD.34.3128,hagen1990aharonov,Semenoff1984PhysRevLett.53.2449,HaldanePhysRevLett.61.2015, Fradkin1987PhysRevLett.57.2967,Chamon2008PhysRevB.77.235431,Jackiw1983}. In the latter case, a dynamical external gauge field induces zero modes of massless planar fermions and vacuum charge with abnormal parity. The Index of the corresponding Dirac operator follows \eqref{eq: pseudocharge} and acquires non-zero values proportional to the strength of the gauge field. 
Hence, the present results provide, for graphene, a measurable realization of these QED effects with the topological content of the gauge field now replaced by vacancies with properly chosen boundary conditions. Furthermore, our findings  display a coherent description of existing measurements \cite{mao2016realization,Ovdat2017} and provide additional predictions that can be tested with an appropriate experimental control on multi-vacancy configurations. \Omrie{Several aspects of these features may not be realized in an experimental setup. Due to noise and interactions vacancies will only be correlated up to some finite screening length. Within this regime, interactions may also result in a broadening and delocalizing of charge around the vacancies especially if these are tightly packed. It would be interesting to study the extent of this effect in the framework of an interacting model such as the Hubbard model.}

Including spin degrees of freedom in the Dirac picture and connecting with Lieb's theorem \cite{Lieb1989} may enrich the picture presented here by associating to a vacancy the quantum dynamics of a localized vacuum spin which is  proportional to the Dirac Index. Possible connections to recent observations of vacancy magnetic moments \cite{mao2017quantum,PadmanabhanPhysRevB.93.165403,Ulybyshev2015PhysRevLett.114.246801,Ugeda2010PhysRevLett.104.096804,Valencia2017PhysRevB.96.125431} should be investigated together with a generalisation to other bipartite lattices and to non-isolated vacancies. \sout{The notion of topological switch involves different and somehow unusual algebraic rules, e.g., $3 Q_\blacktriangle + 2 Q_\triangledown = 3 Q_\blacktriangle$ but  $Q_\triangledown + (3 Q_\blacktriangle + 2 Q_\triangledown) =0$, which may have  applications in logic circuit.} 



\paragraph*{Acknowledgements.}

This work was supported by the Israel Science Foundation Grant No.~924/09.


%


\clearpage



\section*{Supplementary material}

# Supplementary Note 1: Relation between vacuum charge and Index $H$

In what follows we show the relation, presented in the main text, between the vacuum charge density $\rho(\boldsymbol{r})$ and the divergence of the vector matrix element $\boldsymbol{\Delta}(\boldsymbol{r}) \equiv \frac{1}{2}\left\langle \boldsymbol{r} \left| \mathrm{tr}\left(i\boldsymbol{\sigma}^i \sigma_3 \frac{1}{H-i0}\right) \right| \boldsymbol{r} \right\rangle$.

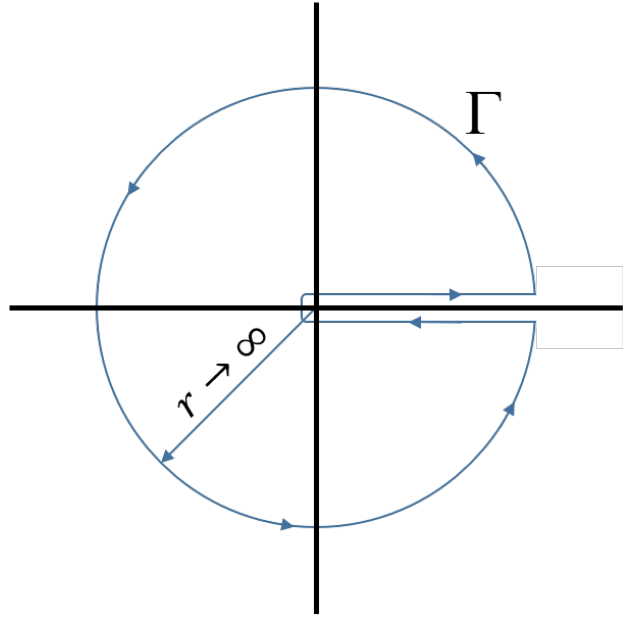

Figure 1: **Integration contour.** Contour of integration performed in (4)

As shown in [1, 2]

$$\rho(\boldsymbol{r}) = \frac{e}{2} \sum_n \mathrm{sign}(E_n) \psi_n^\dagger(\boldsymbol{r}) \psi_n(\boldsymbol{r})$$
$$= M \frac{e}{\pi} \lim_{s \to 0^+} \cos\left(\frac{\pi s}{2}\right) \int_0^\infty d\omega\, \omega^{-s} \frac{1}{M^2 + \omega^2} \Omega\left(\boldsymbol{r}, \sqrt{M^2 + \omega^2}\right) \tag{1}$$

where $e > 0$,

$$\Omega(\boldsymbol{r}, z) \equiv -iz \left\langle \boldsymbol{r} \left| \mathrm{tr}\left(\sigma_3 (H - iz)^{-1}\right) \right| \boldsymbol{r} \right\rangle, \tag{2}$$

$H = -i\boldsymbol{\sigma} \cdot \boldsymbol{\nabla}$ and $M$ is a regularising mass parameter. Using contour integration we can calculate the above integral. Since $\sqrt{M^2 + \omega^2} > 0$, $f(\omega) \equiv \Omega\left(\boldsymbol{r}, \sqrt{M^2 + \omega^2}\right)$ is exponentially decaying for $\omega \to \infty$, has no poles and obeys $f(-\omega) = f(\omega)$. Consider the integral

$$I = \int_0^\infty d\omega\, \omega^{-s} \frac{1}{M^2 + \omega^2} f(\omega). \tag{3}$$

The function $z^{-s} = e^{-s \log z} \equiv re^{i\theta}$ has a branch cut that can be defined on the positive real line with $0 < \theta < 2\pi$. The poles of $(M^2 + \omega^2)$ are at $\omega = \pm i|M|$. Integrating over the contour $\Gamma$ shown in supplementary Fig. 1 and using the residue theorem

$$\oint_\Gamma dz\, z^{-s} \frac{1}{M^2 + z^2} f(z) = \frac{\pi}{|M|} (i|M|)^{-s} f(i|M|)(1 - e^{-i\pi s})$$
$$= (1 - e^{-2\pi i s}) I \qquad (4)$$

thus

$$I = \frac{\pi}{2|M|^{s+1} \cos \frac{\pi s}{2}} \lim_{z \to 0} \Omega(\boldsymbol{r}, z) \qquad (5)$$

and

$$\rho(\boldsymbol{r}) = M \frac{e}{\pi} \lim_{s \to 0^+} \cos\left(\frac{\pi s}{2}\right) \left(\frac{\pi}{2|M|^{s+1} \cos \frac{\pi s}{2}} \lim_{z \to 0} \Omega(\boldsymbol{r}, z)\right)$$
$$= \frac{e}{2} \operatorname{sign} M \lim_{z \to 0} \Omega(\boldsymbol{r}, z). \qquad (6)$$

Using the identity [1, 3]

$$\Omega(\boldsymbol{r}, z) = \frac{1}{2} \boldsymbol{\nabla} \cdot \left\langle \boldsymbol{r} \left| \operatorname{tr}\left(i\boldsymbol{\sigma}^i \sigma_3 \frac{1}{H - iz}\right) \right| \boldsymbol{r} \right\rangle \qquad (7)$$

we obtain the alternative form

$$\rho(\boldsymbol{r}) = \frac{e}{2} \operatorname{sign}(M) \boldsymbol{\nabla} \cdot \boldsymbol{\Delta}(\boldsymbol{r}). \qquad (8)$$

## Supplementary Note 2: Most general boundary condition of an non-penetrable circular wall

In what follows we derive the analogue of the 'mixed boundary condition' for the case of the Dirac Hamiltonian.

Consider the free Dirac Hamiltonian $H = \boldsymbol{\sigma} \cdot \boldsymbol{p}$. The matrix element of the difference $H - H^\dagger$ is a boundary term

$$\langle g| (H - H^\dagger) |f\rangle = \int d\boldsymbol{r}\, g(\boldsymbol{r})^\dagger (-i\boldsymbol{\sigma} \cdot \boldsymbol{\nabla} f(\boldsymbol{r})) - \int d\boldsymbol{r}\, (-i\boldsymbol{\sigma} \cdot \boldsymbol{\nabla} g(\boldsymbol{r}))^\dagger f(\boldsymbol{r})$$
$$= -i \int d\boldsymbol{S} \cdot \left(g(\boldsymbol{r})^\dagger \boldsymbol{\sigma} f(\boldsymbol{r})\right). \qquad (9)$$

In terms of Dirac gamma matrices, $H = \boldsymbol{\sigma} \cdot \boldsymbol{p} = \gamma^0 \gamma^i p^i$, thus

$$\langle \psi | (H - H^\dagger) | \psi \rangle = -i \int dS^i \left( \bar{\psi}(\boldsymbol{r}) \gamma^i \psi(\boldsymbol{r}) \right) \tag{10}$$

which is proportional to the current density. To impose $H = H^\dagger$ we a require a boundary condition on all eigenfunctions of $H$ such that (9) vanishes. The corresponding boundary thus represents a perfect reflector of probability current density. For the case of a circular boundary of radius $R$ around the origin

$$\langle g | (H - H^\dagger) | f \rangle = R \int d\theta \, g(\boldsymbol{r})^\dagger \begin{pmatrix} 0 & e^{-i\theta} \\ e^{i\theta} & 0 \end{pmatrix} f(\boldsymbol{r})$$

$$= R \int d\theta \left( g^{A*}(\boldsymbol{r}) f^B(\boldsymbol{r}) e^{-i\theta} + g^{B*}(\boldsymbol{r}) f^A(\boldsymbol{r}) e^{i\theta} \right) \tag{11}$$

where we used the identity $\hat{r} \cdot \boldsymbol{\sigma} = \begin{pmatrix} 0 & e^{-i\theta} \\ e^{i\theta} & 0 \end{pmatrix}$ and defined $f(\boldsymbol{r}) = ( f^A \; f^B )^T$, $g(\boldsymbol{r}) = ( g^A \; g^B )^T$. The general set of solutions to $H\psi = E\psi$, given in terms of polar coordinates, is spanned by the basis

$$\psi_{k,m,\lambda}(\boldsymbol{r}) = e^{im\theta} \begin{pmatrix} \psi_{k,m,\lambda}^A(r) \\ \lambda i \psi_{k,m,\lambda}^B(r) e^{i\theta} \end{pmatrix} \tag{12}$$

where $k \equiv |E|$, $\lambda \equiv \text{sign } E$ and $m \in \mathbb{Z}$, $j \equiv m + 1/2$ are the orbital and total angular momentum numbers respectively. Consider the eigenfunctions $f_{k,m,\lambda}(\boldsymbol{r}), g_{k',m',\lambda'}(\boldsymbol{r})$. The corresponding boundary term reduces to

$$\langle g | (H - H^\dagger) | f \rangle = iR \left( \lambda g_{k',m',\lambda'}^{A*}(R) f_{k,m,\lambda}^B(R) - \lambda' g_{k',m',\lambda'}^{B*}(R) f_{k,m,\lambda}^A(R) \right) \int_0^{2\pi} d\theta \, e^{i(m-m')\theta}$$

$$= 2\pi i R \delta_{mm'} \left( \lambda g_{k',m',\lambda'}^{A*}(R) f_{k,m,\lambda}^B(R) - \lambda' g_{k',m',\lambda'}^{B*}(R) f_{k,m,\lambda}^A(R) \right). \tag{13}$$

For all eigenfunctions, we take the boundary conditions $\psi_{k,m,\lambda}^A / \psi_{k,m,\lambda}^B = \lambda h_m$ where $h_m$ is some real, energy independent number, then

$$\langle g | (H - H^\dagger) | f \rangle = 2\pi i R \lambda \lambda' \delta_{mm'} (h_{m'} - h_m) g_{k',m',\lambda'}^{B*}(R) f_{k,m,\lambda}^B(R) = 0. \tag{14}$$

## Supplementary Note 3: Boundary conditions in the presence of circular vacancy of type B

The analogue choice of boundary conditions corresponding to a $B$-vacancy are

$$\begin{aligned} \psi_m^A(R) &= 0, \quad m < -1 \\ \psi_m^B(R) &= 0, \quad m \geq -1. \end{aligned} \tag{15}$$

## Supplementary Note 4: The charge density in the presence of a single vacancy

In what follows, we obtain the explicit expression for the vacuum charge density in the case of a single vacancy.

As explained in the main text, the charge density is given by (sign $M \equiv 1$)

$$Q = \int d\boldsymbol{r}\, \rho(\boldsymbol{r}) = -\frac{e}{2}\, \text{Index}\, H. \tag{16}$$

where,

$$\text{Index}\, H = \lim_{z \to 0} \text{Tr}\left(\frac{z}{H^B + z} - \frac{z}{H^A + z}\right) \tag{17}$$

with $H^B = D^\dagger D$, $H^A = DD^\dagger$. In what follows, we use relation (17) to obtain Index $H$ and consequently $\rho(\boldsymbol{r})$.

Assuming $E > 0$, $H\psi = E\psi$ is given in terms of polar coordinates by

$$-\psi_m^{A\prime}(r) + \frac{m}{r}\psi_m^A(r) = E\psi_m^B \tag{18a}$$

$$\psi_m^{B\prime}(r) + \frac{(m+1)}{r}\psi_m^B(r) = E\psi_m^A. \tag{18b}$$

where

$$\psi(\boldsymbol{r}) = \sum_{m=-\infty}^{\infty} e^{im\theta} \begin{pmatrix} \psi_m^A(r) \\ i\psi_m^B(r)e^{i\theta} \end{pmatrix}. \tag{19}$$

The set of first order equations (18) can be decoupled into two independent second order equations

$$H^A \psi^A = E^2 \psi^A \tag{20a}$$

$$H^B \psi^B = E^2 \psi^B \tag{20b}$$

where both $H^A, H^B$ formally equal to $-\nabla^2$. Assuming the boundary conditions corresponding to an $A$-vacancy (see main text), supplementary Eqs. (20) read

$$\left(-\partial_r^2 - \frac{1}{r}\partial_r + \frac{m^2}{r^2}\right)\psi_m^A(r) = E^2 \psi_m^A(r), \quad \begin{cases} \psi_m^A(R) = 0 & m \leq 0 \\ \psi_m^{A\prime}(R)/\psi_m^A(R) = \frac{m}{R} & m > 0 \end{cases} \tag{21a}$$

$$\left(-\partial_r^2 - \frac{1}{r}\partial_r + \frac{(m+1)^2}{r^2}\right)\psi_m^B(r) = E^2 \psi_m^B(r), \quad \begin{cases} \psi_m^{B\prime}(R)/\psi_m^B(R) = -\frac{m+1}{R} & m \leq 0 \\ \psi_m^B = 0 & m > 0. \end{cases} \tag{21b}$$

The Dirichlet conditions on $\psi_m^A, \psi_m^B$ correspond to the chiral boundary conditions proposed in the main text to describe the physics of the vacancy. Combined with the Dirac equation (18), these constrain the additional mixed boundary conditions appearing in (21). To make (21) more symmetrical we transform

5

$m \to -m-1$ in the second equation and redefine $\psi^B_{-m-1} \to \psi^B_m$ such that

$$\left(-\partial_r^2 - \frac{1}{r}\partial_r + \frac{m^2}{r^2}\right)\psi^A_m(r) = E^2 \psi^A_m(r), \quad \begin{cases} \psi^A_m(R) = 0 & m \leq 0 \\ \psi^{A\prime}_m(R)/\psi^A_m(R) = \frac{m}{R} & m > 0 \end{cases} \quad (22a)$$

$$\left(-\partial_r^2 - \frac{1}{r}\partial_r + \frac{m^2}{r^2}\right)\psi^B_m(r) = E^2 \psi^B_m(r), \quad \begin{cases} \psi^B_m = 0 & m < -1 \\ \psi^{B\prime}_m(R)/\psi^B_m(R) = \frac{m}{R} & m \geq -1. \end{cases} \quad (22b)$$

Define $G^{A/B}(z) \equiv \frac{1}{H^{A/B}+z}$. In position space,

$$\int d\boldsymbol{r}'' \langle \boldsymbol{r}| H^{A/B} + z |\boldsymbol{r}''\rangle \langle \boldsymbol{r}''| G^{A/B} |\boldsymbol{r}'\rangle = \frac{1}{r}\delta(r-r')\delta(\theta-\theta'), \quad (23)$$

or, equivalently,

$$\left(-\partial_r^2 - \frac{1}{r}\partial_r - \frac{1}{r^2}\partial_\theta^2 + z\right) G^{A/B}(\boldsymbol{r}, \boldsymbol{r}') = \frac{1}{r}\delta(r-r')\frac{1}{2\pi}\sum_{m=-\infty}^{\infty} e^{im(\theta-\theta')}, \quad (24)$$

where we used the identity

$$\delta(\theta-\theta') = \frac{1}{2\pi}\sum_{m=-\infty}^{\infty} e^{im(\theta-\theta')}. \quad (25)$$

After insertion of the following expansion

$$G^{A/B}(\boldsymbol{r}, \boldsymbol{r}') = \frac{1}{2\pi}\sum_{m=-\infty}^{\infty} G^{A/B}_m(r, r') e^{im(\theta-\theta')}, \quad (26)$$

supplementary Eq. (24) reduces to the set

$$\left(-\partial_r^2 - \frac{1}{r}\partial_r + \frac{m^2}{r^2} + z\right) G^{A/B}_m(r, r') = \frac{1}{r}\delta(r-r'). \quad (27)$$

Although $G^{A/B}_m(r, r')$ obey the same (trivial) equation, they are constrained to different sets of boundary conditions corresponding to (22)

$$\left(-\partial_r^2 - \frac{1}{r}\partial_r + \frac{m^2}{r^2} + z\right) G^A_m(r, r') = \frac{1}{r}\delta(r-r'), \quad \begin{cases} G^A_m(R, r') = 0 & m \leq 0 \\ \partial_r G^A_m(R, r')/G^A_m(R, r') = \frac{m}{R} & m > 0 \end{cases} \quad (28a)$$

$$\left(-\partial_r^2 - \frac{1}{r}\partial_r + \frac{m^2}{r^2} + z\right) G^B_m(r, r') = \frac{1}{r}\delta(r-r'), \quad \begin{cases} G^B_m(R, r') = 0 & m < -1 \\ \partial_r G^B_m(R, r')/G^B_m(R, r') = \frac{m}{R} & m \geq -1. \end{cases} \quad (28b)$$

Table 1: **Boundary condition *A*-vacancy.** Boundary conditions for an *A*-vacancy imposed on the radial components of the resolvent operators $G^A$, $G^B$. The conditions are symmetrical $\forall m \neq 0, -1$ ($\forall j \neq \pm 1/2$).

| $m$ | $G_m^A$ | $G_m^B$ |
|---|---|---|
| $\leq -2$ | Dirichlet | Dirichlet |
| $-1$ | Dirichlet | Mixed |
| $0$ | Dirichlet | Mixed |
| $\geq 1$ | Mixed | Mixed |

The boundary conditions in (28) are symmetrical with respect to the label $A, B$ for all $m \neq 0, -1$ ($j \neq \pm 1/2$) as shown in Tab. 1. We further require that $G_m^{A/B}(r, r')$ decay for $r, r' \to \infty$.

The solutions of (28) are given by

$$G_m^{A/B} = I_m\left(\sqrt{z}r_<\right) K_m\left(\sqrt{z}r_>\right) + \Gamma_m^{A/B} K_m\left(\sqrt{z}r\right) K_m\left(\sqrt{z}r'\right) \tag{29}$$

where $I_n(x), K_n(x)$ are the modified Bessel functions of the first and second kind, $r_< \equiv \min(r, r')$, $r_> \equiv \max(r, r')$ and $\Gamma_m^{A/B}$ are coefficients to be determined by boundary conditions. The first term in (29) is a private solution of the non-homogeneous differential equation in (28) [4]. The second term is a solution of the corresponding homogeneous equation and is required so that (29) obeys the necessary boundary conditions. Imposing these conditions and utilizing the symmetry expressed in Tab. 1 gives

$$\Gamma_m^A = \Gamma_m^B \quad \forall m \neq 0, -1. \tag{30}$$

In addition,

$$\Gamma_0^A = -\Gamma_{-1}^B = -I_0\left(\sqrt{z}R\right)/K_0\left(\sqrt{z}R\right) \tag{31a}$$

$$\Gamma_0^B = -\Gamma_{-1}^A = I_1\left(\sqrt{z}R\right)/K_1\left(\sqrt{z}R\right), \tag{31b}$$

and

$$\Gamma_0^B - \Gamma_0^A = \Gamma_{-1}^B - \Gamma_{-1}^A = \frac{1}{\sqrt{z}R K_0(\sqrt{z}R) K_1(\sqrt{z}R)}. \tag{32}$$

Finally we arrive to

$$\begin{aligned}
\text{Index } H &= \lim_{z \to 0} z \, \text{Tr} \left( \frac{1}{H^B + z} - \frac{1}{H^A + z} \right) \\
&= \lim_{z \to 0} z \, \text{tr} \left( G^B - G^A \right) \\
&= \frac{z}{2\pi} \int d\boldsymbol{r} \sum_m \left( G_m^B(r,r) - G_m^A(r,r) \right) \\
&= \frac{z}{2\pi} \lim_{z \to 0} \int d\boldsymbol{r} \left[ \left( \Gamma_0^B - \Gamma_0^A \right) K_0 \left( \sqrt{z}r \right)^2 + \left( \Gamma_{-1}^B - \Gamma_{-1}^A \right) K_{-1} \left( \sqrt{z}r \right)^2 \right] \\
&= \frac{z}{2\pi} \lim_{z \to 0} \left( \Gamma_0^B - \Gamma_0^A \right) \int d\boldsymbol{r} \left( K_0 \left( \sqrt{z}r \right)^2 + K_{-1} \left( \sqrt{z}r \right)^2 \right).
\end{aligned} \tag{33}$$

Inserting (32) and using the identity

$$K_0 \left( \sqrt{z}r \right)^2 + K_1 \left( \sqrt{z}r \right)^2 = -\frac{1}{\sqrt{z}} \boldsymbol{\nabla} \cdot \left( K_0(\sqrt{z}r) K_1(\sqrt{z}r) \hat{r} \right), \tag{34}$$

we obtain

$$\text{Index } H = -\frac{1}{2\pi R} \lim_{z \to 0} \int d\boldsymbol{r} \, \boldsymbol{\nabla} \cdot \left( \frac{K_0(\sqrt{z}r) K_1(\sqrt{z}r)}{K_0(\sqrt{z}R) K_1(\sqrt{z}R)} \hat{r} \right). \tag{35}$$

Integrating (35) in the region $R < r < \infty$, $0 < \theta < 2\pi$ and inserting into (16) gives

$$Q = -\frac{e}{2} \text{Index } H = -\frac{e}{2} \cdot \left( \lim_{z \to 0} 1 \right) = -\frac{e}{2}. \tag{36}$$

The charge density $\rho(\boldsymbol{r})$ can be read off the integrand in (35)

$$\rho(\boldsymbol{r}) = -\frac{e}{4\pi R} \boldsymbol{\nabla} \cdot \left( \frac{K_0(\sqrt{z}r) K_1(\sqrt{z}r)}{K_0(\sqrt{z}R) K_1(\sqrt{z}R)} \hat{r} \right). \tag{37}$$

In the limit of a pointlike vacancy, $R \to 0$, $\rho(\boldsymbol{r})$ vanishes $\forall r \neq 0$. Since $\int d\boldsymbol{r} \, \rho(\boldsymbol{r}) = -e/2$, independent of $R$, $\rho(\boldsymbol{r})$ can be represented by the $\delta$-function distribution

$$\lim_{R \to 0} \rho(\boldsymbol{r}) = -\frac{1}{2\pi} \boldsymbol{\nabla} \cdot \left( \frac{e/2}{r} \hat{r} \right). \tag{38}$$

For finite $R$, $\rho(\boldsymbol{r})$ can be approximated from (37) with an arbitrarily small finite value of $z$ acting as an IR cutoff [1]. For $r\sqrt{z} \gg 1$, $\rho(r)/\rho(R) \approx e^{-2\sqrt{z}r}$ and for $r\sqrt{z} \ll 1$, $R\sqrt{z} \ll 1$, $\rho(r)/\rho(R) \approx R^2/r^2$. Thus, the charge density decays close to the vacancy as $\sim 1/r^2$ and decays exponentially far from the vacancy (see

---

[1] Note that it is not allowed to move the $\lim_{z \to 0}$ through the integral in (35) since

$$\lim_{z \to 0} \lim_{r \to \infty} \frac{2\pi r}{2\pi R} \frac{K_0(\sqrt{z}r) K_1(\sqrt{z}r)}{K_0(\sqrt{z}R) K_1(\sqrt{z}R)} = 0 \neq \lim_{r \to \infty} \lim_{z \to 0} \frac{2\pi r}{2\pi R} \frac{K_0(\sqrt{z}r) K_1(\sqrt{z}r)}{K_0(\sqrt{z}R) K_1(\sqrt{z}R)} = 1$$

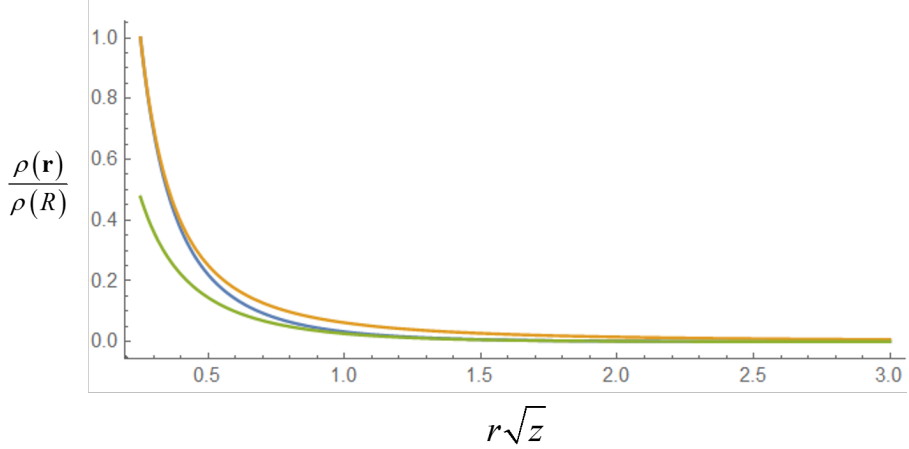

Figure 2: **Single vacancy charge density.** Blue: Characteristic behaviour of $\rho(\boldsymbol{r})/\rho(R)$ in (37) as a function of $x \equiv r\sqrt{z}$ with $y \equiv R\sqrt{z} = 0.25$. Orange: the function $y^2/x^2$. Green: The function $\pi y^2 e^{-2x}/x$

Table 2: **Boundary condition $A$-vacancy.** Boundary conditions for a $B$-vacancy imposed on the radial components of the resolvent operators $G^A$, $G^B$. The conditions are symmetrical $\forall m \neq 0, -1$ ($\forall j \neq \pm 1/2$) and are the same those presented in Tab. 1 up to the exchange of columns $A \leftrightarrow B$.

| $m$ | $G_m^A$ | $G_m^B$ |
|---|---|---|
| $\leq -2$ | Dirichlet | Dirichlet |
| $-1$ | Mixed | Dirichlet |
| $0$ | Mixed | Dirichlet |
| $\geq 1$ | Mixed | Mixed |

supplementary Fig. 2).

In the case of a $B$-vacancy, corresponding to boundary conditions (15), the analogue equations of $G_m^{A/B}(r, r')$ are

$$\left(-\partial_r^2 - \frac{1}{r}\partial_r + \frac{m^2}{r^2} + z\right) G_m^A(r, r') = \frac{1}{r}\delta(r - r'), \quad \begin{cases} G_m^A(R, r') = 0 & m < -1 \\ G_m^{A\prime}(R)/G_m^A(R) = \frac{m}{R} & m \geq -1 \end{cases} \quad (39a)$$

$$\left(-\partial_r^2 - \frac{1}{r}\partial_r + \frac{m^2}{r^2} + z\right) G_m^B(r, r') = \frac{1}{r}\delta(r - r'), \quad \begin{cases} G_m^B(R, r') = 0 & m \leq 0 \\ G_m^{B\prime}(R, r')/G_m^B(R, r') = \frac{m}{R} & m > 0. \end{cases} \quad (39b)$$

The boundary conditions in (39), can be summarized in Tab. 2 which is identical to Tab. 1 up to the exchange of columns $A \leftrightarrow B$. Thus, from (17) it is apparent that the calculation of the charge density will follow as in the case of the $A$-vacancy but with an opposite sign.

9

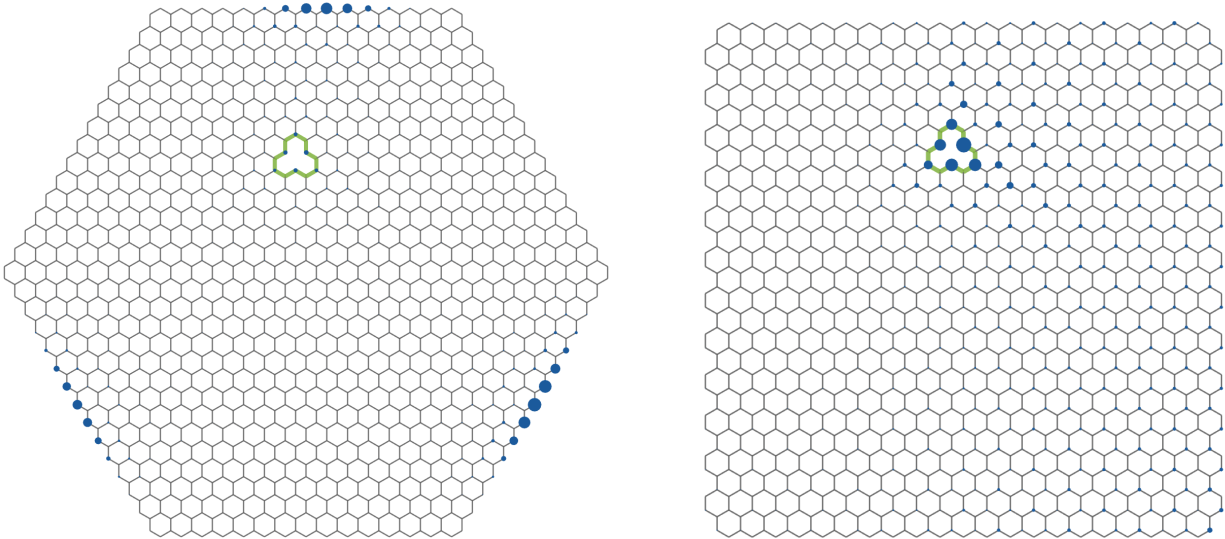

Figure 3: **Boundary effects on charge distribution. a.** Zig-zag boundary on a hexagonal sheet. A large portion of the charge is distributed on the edges corresponding to the majority sublattice. **b.** Periodic (cylindrical) boundary conditions on a square sheet. The periodic boundary is on the horizontal (zig-zag) edges. As a result, there is no accumulation of charge on the 'zigzag edge' in this case.

## Supplementary Note 5: Armchair, zigzag and periodic boundary conditions

We present in supplementary Fig. 3 the effects of using a zigzag edge as opposed to an armchair edge in the numerical diagonalisation of the (finite) tight binding Hamiltonian. In the presence of an open zigzag edge (supplementary Fig. 3a) some of the charge accumulates at the corresponding boundary. This effect can be removed by imposing periodic boundary conditions on this direction (supplementary Fig. 3b) or using armchair boundary conditions in all directions (as in main text).

## Supplementary Note 6: The charge density in the presence of a multiple vacancies

In what follows we present the formalism of low energy scattering theory in which we obtain a closed form expression of the charge density for the case of a multi-vacancy configuration. Using this expression we were able to generate Figs. 2b, 3b of the main text.

**General scattering theory**

Consider a Dirac particle in a $2+1$ dimensional plane with one puncture at the origin. The Hamiltonian is $H = \boldsymbol{\sigma} \cdot \boldsymbol{p}$. The general $E > 0$ solution, written in terms of polar coordinates is

$$\psi(\boldsymbol{r}) = \sum_{m \in \mathbb{Z}} \frac{i^m}{2} e^{im\theta} e^{-im\theta_k} \left( \begin{pmatrix} H_m^{(2)}(kr) \\ ie^{i\theta} H_{m+1}^{(2)}(kr) \end{pmatrix} + e^{2i\delta_m} \begin{pmatrix} H_m^{(1)}(kr) \\ ie^{i\theta} H_{m+1}^{(1)}(kr) \end{pmatrix} \right) \tag{40}$$

where $\boldsymbol{k} = |E|(\cos\theta_k, \sin\theta_k)$, $j = m + 1/2$ is the total angular momentum, $\delta_m(k)$ is the scattering phase shift and $H_m^{(1)}(x), H_m^{(2)}(x)$ are the Hankel functions of the first and second kind. We can expand $\psi$ to the form of an incoming plane wave and outgoing scattered radial wave

$$\begin{aligned}
\psi(\boldsymbol{r}) &= \sum_{m \in \mathbb{Z}} \frac{i^m}{2} e^{im\theta} e^{-im\theta_k} \left( \begin{pmatrix} H_m^{(2)}(kr) \\ ie^{i\theta} H_{m+1}^{(2)}(kr) \end{pmatrix} + \begin{pmatrix} H_m^{(1)}(kr) \\ ie^{i\theta} H_{m+1}^{(1)}(kr) \end{pmatrix} \right) \\
&+ \sum_{m \in \mathbb{Z}} \frac{i^m}{2} e^{im\theta} e^{-im\theta_k} \left( e^{2i\delta_m} - 1 \right) \begin{pmatrix} H_m^{(1)}(kr) \\ ie^{i\theta} H_{m+1}^{(1)}(kr) \end{pmatrix} \\
&= e^{i\boldsymbol{k} \cdot \boldsymbol{r}} \begin{pmatrix} 1 \\ e^{i\theta_k} \end{pmatrix} + \sum_{m \in \mathbb{Z}} i^m e^{im(\theta-\theta_k)} f_m(k) k^{1/2} \begin{pmatrix} H_m^{(1)}(kr) \\ ie^{i\theta} H_{m+1}^{(1)}(kr) \end{pmatrix},
\end{aligned} \tag{41}$$

such that,

$$\psi(\boldsymbol{r}) \stackrel{r \to \infty}{\approx} e^{i\boldsymbol{k} \cdot \boldsymbol{r}} \begin{pmatrix} 1 \\ e^{i\theta_k} \end{pmatrix} + f(\boldsymbol{k}, \theta) \begin{pmatrix} 1 \\ e^{i\theta} \end{pmatrix} \frac{e^{ikr}}{\sqrt{r}}, \tag{42}$$

where we used the identities

$$e^{i\boldsymbol{k} \cdot \boldsymbol{r}} = \sum_{m \in \mathbb{Z}} \frac{i^m}{2} e^{im(\theta-\theta_k)} \left( H_m^{(2)}(kr) + H_m^{(1)}(kr) \right) \tag{43a}$$

$$H_m^{(1)}(kr) = \frac{(1-i)i^{-m}}{\sqrt{\pi k r}} e^{ikr} + \mathcal{O}\left(r^{-3/2}\right), \tag{43b}$$

and defined,

$$f_m(k) \equiv \left( e^{2i\delta_m} - 1 \right) / 2\sqrt{k} \tag{44a}$$

$$f(\boldsymbol{k}, \theta) \equiv \sqrt{\frac{1}{\pi}} (1-i) \sum_{m \in \mathbb{Z}} e^{im(\theta-\theta_k)} f_m(k), \tag{44b}$$

$f_m$ being the scattering amplitude of the partial wave associated with angular momentum $m$. For a reason that will become clear later, we would like to rearrange the sum (41) into pairs of $m, -m-1$ modes

11

(equivalent to $j = \pm(m + 1/2)$)

$$\psi = e^{i\boldsymbol{k}\cdot\boldsymbol{r}}\begin{pmatrix}1\\e^{i\theta_k}\end{pmatrix} + \sum_{m\in\mathbb{Z}} i^m e^{im(\theta-\theta_k)} f_m(k) k^{1/2} \begin{pmatrix} H_m^{(1)}(kr) \\ ie^{i\theta} H_{m+1}^{(1)}(kr) \end{pmatrix}$$

$$= \left(e^{i\boldsymbol{k}\cdot\boldsymbol{r}} + \frac{4}{ik^{1/2}} \sum_{m\geq 0} G_m^f(\boldsymbol{r};k) F_m(k) U^m \right) \begin{pmatrix}1\\e^{i\theta_k}\end{pmatrix} \tag{45}$$

where we used the identity $H_{-m}^{(1)}(x) = e^{i\pi m} H_m^{(1)}(x)$, and

$$G_m^f(\boldsymbol{r};k) \equiv \frac{i^{m+1}k}{4} \begin{pmatrix} H_m^{(1)}(kr) & ie^{-i\theta} H_{m+1}^{(1)}(kr) \\ ie^{i\theta} H_{m+1}^{(1)}(kr) & H_m^{(1)}(kr) \end{pmatrix}, \tag{46a}$$

$$U \equiv \begin{pmatrix} e^{i(\theta-\theta_k)} & 0 \\ 0 & e^{-i(\theta-\theta_k)} \end{pmatrix}, \tag{46b}$$

$$F_m(k) \equiv \begin{pmatrix} f_m(k) & 0 \\ 0 & f_{-m-1}(k) \end{pmatrix}. \tag{46c}$$

The form presented in (45) shows that the amplitude of $\psi$ is a sum of two contributions: an incoming plane wave and a scattered wave given by a sum over angular momentum contributions $G_m^f F_m U^m$. The physics of the scatterer is completely encoded in $F_m(k)$. The amplitudes $G_m^f$ are intrinsic characteristics of the free system.

In the low energy regime, $kR \ll 1$, $R$ being the range of the potential or vacancy in our case, the scattering amplitude in the lowest angular momentum channels are generally the most dominant and all higher partial waves can be neglected giving

$$\psi = \left(e^{i\boldsymbol{k}\cdot\boldsymbol{r}} + \frac{4}{ik^{1/2}} G_0^f(\boldsymbol{r};k) F_0(k)\right) \begin{pmatrix}1\\e^{i\theta_k}\end{pmatrix}. \tag{47}$$

Hereafter, we refer strictly to $G_0^f$, $F_0$ and neglect their subscript for brevity. In the case of an arbitrary incoming wave packet

$$\Phi(\boldsymbol{r}) = \int d\theta_k \Phi(\theta_k) \begin{pmatrix}1\\e^{i\theta_k}\end{pmatrix} e^{i\boldsymbol{k}\cdot\boldsymbol{r}}, \tag{48}$$

composed out of plane waves with $|k| = E$, an immediate generalization of (47) gives

$$\psi = \Phi(\boldsymbol{r}) + \frac{4}{ik^{1/2}} G^f(\boldsymbol{r};k) F(k) \Phi(0). \tag{49}$$

Note that

$$G^f(\boldsymbol{r};k) = (-i\boldsymbol{\sigma}\cdot\boldsymbol{\nabla} + k)\left(\frac{i}{4} H_0^{(1)}(kr)\right) \tag{50}$$

12

is the outgoing Green's function of $H - k$ on the free plane (i.e. no scatterers),

$$\begin{aligned}(H - k) G^f (\boldsymbol{r}; k) &= (-i\boldsymbol{\sigma} \cdot \boldsymbol{\nabla} - k)(-i\boldsymbol{\sigma} \cdot \boldsymbol{\nabla} + k) \left(\frac{i}{4} H_0^{(1)}(k\boldsymbol{r})\right) \\ &= -\left(\nabla^2 + k^2\right) \left(\frac{i}{4} H_0^{(1)}(k\boldsymbol{r})\right) \\ &= \delta(\boldsymbol{r}) \mathbf{1}_{2\times 2}.\end{aligned} \quad (51)$$

This relation reflects the fact that, in the limit $kR \ll 1$, the scattered wave corresponds a wave radiated by a point source with an amplitude given by $F_0(k) \Phi(0)$.

In what follows we show explicitly that indeed only the $m = 0, -1$ (s-wave) scattering amplitudes are important for $kR \ll 1$, $R$ being that range of the vacancy.

**The scattering amplitudes $f_m$**

The boundary conditions corresponding to a vacancy from sublattice $A, B$ given in the main text and in (15) are

$$\psi^A_{m\leq 0}(R) = \psi^B_{m>0}(R) = 0 \quad (52a)$$

and

$$\psi^A_{m<-1}(R) = \psi^B_{m\geq -1}(R) = 0 \quad (52b)$$

respectively. Applying (52a) to the general wave function solution (40) gives

$$e^{2i\delta^A_m} = \begin{cases} -\dfrac{H^{(2)}_{m+1}(kR)}{H^{(1)}_{m+1}(kR)} & m > 0 \\ -\dfrac{H^{(2)}_m(kR)}{H^{(1)}_m(kR)} & m \leq 0 \end{cases} \quad (53)$$

which, from (44a), corresponds to

$$f^A_m(k) = -\frac{1}{\sqrt{k}} \begin{cases} \dfrac{J_{m+1}(kR)}{H^{(1)}_{m+1}(kR)} & m > 0 \\ \dfrac{J_m(kR)}{H^{(1)}_m(kR)} & m \leq 0 \end{cases} \quad (54a)$$

$$\overset{kR\ll 1}{\sim} \frac{1}{\sqrt{k}} \begin{cases} (kR)^{2(m+1)} & m > 0 \\ \dfrac{1}{\log(kR)} & m = 0 \\ (kR)^{-2m} & m < 0 \end{cases} \quad (54b)$$

where $J_m(x)$ is the Bessel function. Note that

$$f^A_m = f^A_{-m-1} \quad, \forall m \neq 0, -1. \quad (55)$$

13

Applying (52b) to (40) gives

$$e^{2i\delta_m^B} = \begin{cases} -\frac{H_{m+1}^{(2)}(kR)}{H_{m+1}^{(1)}(kR)} & m \geq -1 \\ -\frac{H_m^{(2)}(kR)}{H_m^{(1)}(kR)} & m < -1 \end{cases} \tag{56}$$

which corresponds to

$$f_m^B(k) = -\frac{1}{\sqrt{k}} \begin{cases} \frac{J_{m+1}(kR)}{H_{m+1}^{(1)}(kR)} & m \geq -1 \\ \frac{J_m(kR)}{H_m^{(1)}(kR)} & m < -1 \end{cases} \tag{57}$$

Note that $f_m^B = f_m^A \,\forall m \neq 0, -1$ and that

$$f_0^B = f_{-1}^A, \; f_{-1}^B = f_0^A. \tag{58}$$

From (54b) and (58), it is apparent that for $kR \ll 1$, all the partial wave scattering amplitudes vanish except $f_0^A = f_{-1}^B$ which diverge. Thus, the most dominant contributions to the scattering amplitude arrives from the $j = \pm 1/2$ (s-wave) channel.

**Vacuum charge density for a general configuration of multiple vacancies**

Utilizing the formalism above, we obtain a closed form expression for the charge density $\rho(\mathbf{r})$ in the framework of the continuous Dirac model. Although not illuminating at first sight, this expression allows to plot figures 2b, 3b of the main text.

The charge density can be written in the form (6) (sign $M \equiv 1$)

$$\rho(\mathbf{r}) = \frac{e}{2} \lim_{z \to 0} \Omega(\mathbf{r}, z) \tag{59}$$

where $\Omega(\mathbf{r}, z) = -iz \langle \mathbf{r}| \text{tr}\left(\sigma_3 (H - iz)^{-1}\right) |\mathbf{r}\rangle$. We would like to obtain the matrix element $\Omega(\mathbf{r}, z)$ for a general vacancy configuration. In the low energy regime, $kR \ll 1$, and in the presence of a single vacancy the solution of the Dirac equation is given by

$$\psi(\mathbf{r}) = \Phi(\mathbf{r}) + \frac{4}{ik^{1/2}} G^f(\mathbf{r}; k) F(k) \Phi(0) \tag{60}$$

as shown in supplementary Eq. (49). The physical meaning of this expression is that an incoming wave packet $\Phi(\mathbf{r})$ is scattered as a point source radial wave with amplitude $F(k)\Phi(0)$. In the case of a general vacancy configuration $\mathbf{r}_{iA}$, $\mathbf{r}_{iB}$ expression (60) can be generalised to [5]

$$\psi(\mathbf{r}) = \Phi(\mathbf{r}) + \frac{4}{ik^{1/2}} \sum_{i=1}^{N_A} G^f(\mathbf{r} - \mathbf{r}_{iA}; k) F_A(k) \psi_{i,A} + \frac{4}{ik^{1/2}} \sum_{i=1}^{N_B} G^f(\mathbf{r} - \mathbf{r}_{iB}; k) F_B(k) \psi_{i,B} \tag{61}$$

14

where $F_{A,B}$ correspond to the scattering amplitudes of vacancies $A, B$ respectively and

$$\psi_{i,A} = \Phi(\boldsymbol{r}_{iA}) + \frac{4}{ik^{1/2}} \sum_{\substack{j=1 \\ j \neq i}}^{N_A} G^f(\boldsymbol{r}_{iA} - \boldsymbol{r}_{jA}; k) F_A(k) \psi_{j,A}$$

$$+ \frac{4}{ik^{1/2}} \sum_{j=1}^{N_B} G^f(\boldsymbol{r}_{iA} - \boldsymbol{r}_{jB}; k) F_B(k) \psi_{j,B} \quad (62a)$$

$$\psi_{i,B} = \Phi(\boldsymbol{r}_{iB}) + \frac{4}{ik^{1/2}} \sum_{j=1}^{N_A} G^f(\boldsymbol{r}_{iB} - \boldsymbol{r}_{jA}; k) F_A(k) \psi_{j,A}$$

$$+ \frac{4}{ik^{1/2}} \sum_{\substack{j=1 \\ j \neq i}}^{N_B} G^f(\boldsymbol{r}_{iB} - \boldsymbol{r}_{jB}; k) F_{0,B}(k) \psi_{j,B}. \quad (62b)$$

Supplementary Eqs. (61), (62) simply reflect the fact that the amplitude of $\psi$ is the amplitude of the incoming wave and the amplitude of point source radial waves scattered from each vacancy with scattering amplitudes $F_A(k), F_B(k)$. Coefficients $\psi_{i,A}, \psi_{i,B}$, as given in (62), represent the amplitude at each vacancy point corresponding to the contributions of the incoming wave and all the scattered waves from the other vacancies.

Consider the matrix element $G(\boldsymbol{r}, \boldsymbol{r}'; iz) = \left\langle \boldsymbol{r} \left| (H - iz)^{-1} \right| \boldsymbol{r}' \right\rangle$. It is the Green's function of $H - iz$, that is, it is the response of the system at $\boldsymbol{r}$ to a point source of wave function located at $\boldsymbol{r}'$. Since in (59) we are only interested in the low energy limit $z \to 0$ limit, the expression for $G(\boldsymbol{r}, \boldsymbol{r}'; iz)$ can be obtained from setting $\Phi = G^f(\boldsymbol{r} - \boldsymbol{r}'; k)$ in (61), (62), with $G^f(\boldsymbol{r} - \boldsymbol{r}'; k)$ given in (46a) and $k = iz$

$$G(\boldsymbol{r}, \boldsymbol{r}'; k) = G^f(\boldsymbol{r} - \boldsymbol{r}'; k) + \frac{4}{ik^{1/2}} \sum_{i=1}^{N_A} G^f(\boldsymbol{r} - \boldsymbol{r}_{iA}; k) F_A(k) G_{i,A}$$

$$+ \frac{4}{ik^{1/2}} \sum_{i=1}^{N_B} G^f(\boldsymbol{r} - \boldsymbol{r}_{iB}; k) F_B(k) G_{i,B} \quad (63)$$

and

$$G_{i,A} = G^f\left(\bm{r}_{iA} - \bm{r}'; k\right) + \frac{4}{ik^{1/2}} \sum_{j=1,\,j\neq i}^{N_A} G^f\left(\bm{r}_{iA} - \bm{r}_{jA}; k\right) F_A(k) G_{j,A}$$

$$+ \frac{4}{ik^{1/2}} \sum_{j=1}^{N_B} G^f\left(\bm{r}_{iB} - \bm{r}_{jB}; k\right) F_B(k) G_{j,B} \tag{64a}$$

$$\psi_{i,B} = G^f\left(\bm{r}_{iB} - \bm{r}'; k\right) + \frac{4}{ik^{1/2}} \sum_{j=1}^{N_A} G^f\left(\bm{r}_{iA} - \bm{r}_{jA}; k\right) F_A(k) G_{j,A}$$

$$+ \frac{4}{ik^{1/2}} \sum_{j=1,\,j\neq i}^{N_B} G^f\left(\bm{r}_{iB} - \bm{r}_{jB}; k\right) F_B(k) G_{j,B}. \tag{64b}$$

By solving the linear system (64) we can directly obtain

$$\Omega(\bm{r}, z) = -iz \lim_{\bm{r}' \to \bm{r}} \operatorname{tr}\left(\sigma_3 G(\bm{r}, \bm{r}'; iz)\right) \tag{65}$$

and consequently (59). In Figs. 2b, 3b of the main text we choose $zR = 0.4 \cdot 10^{-6}$

**Vacuum charge density for a single $A$-vacancy using scattering theory**

Utilizing the formalism above, we would like to obtain the vacuum charge density in the presence of a single vacancy. To that purpose we use identity (6) $(\operatorname{sign} M \equiv 1)$

$$\rho(\bm{r}) = \frac{e}{2} \lim_{z \to 0} \Omega(\bm{r}, z) \tag{66}$$

and obtain an explicit expression for the matrix element $\Omega(\bm{r}, z) = -iz \left\langle \bm{r} \left| \operatorname{tr}\left(\sigma_3 (H - iz)^{-1}\right) \right| \bm{r} \right\rangle$.

Consider the matrix element $G(\bm{r}, \bm{r}'; iz) = \left\langle \bm{r} \left| (H - iz)^{-1} \right| \bm{r}' \right\rangle$. It is the Green's function of $H - iz$, that is, it is the response of the system at $\bm{r}$ to a point source of wave function located at $\bm{r}'$. Since in (66) we are only interested in the low energy limit $z \to 0$ the expression for $G(\bm{r}, \bm{r}'; iz)$ can be obtained from setting $\Phi = G^f(\bm{r} - \bm{r}'; k)$ in (49), with $G^f(\bm{r} - \bm{r}'; k)$ given in (46a) and $k = iz$

$$G(\bm{r}, \bm{r}'; k) = G^f(\bm{r} - \bm{r}'; k) + G^f(\bm{r}; k) F(k) G^f(-\bm{r}'; k) \tag{67}$$

and thus

$$\Omega(\bm{r}, z) = -iz \lim_{\bm{r}' \to \bm{r}} \operatorname{tr}\left(\sigma_3 G(\bm{r}, \bm{r}'; iz)\right). \tag{68}$$

16

Using properties of Bessel functions [6], expressions (50) and (54a) can be written as

$$G^f(\boldsymbol{r}; iz) = (-i\boldsymbol{\sigma} \cdot \boldsymbol{\nabla} + iz)\left(\frac{1}{2\pi}K_0(zr)\right) \tag{69}$$

and

$$f_{B,-1} = f_{A,0} = \frac{2\pi i}{z}\frac{I_0(zR)}{K_0(zR)} \tag{70a}$$

$$f_{B,0} = f_{A,-1} = -\frac{2\pi i}{z}\frac{I_1(zR)}{K_1(zR)} \tag{70b}$$

where $I_n(x), K_n(x)$ are the modified Bessel functions of the first and second kind. Using (46c), (69), (70) we can directly obtain (68). Note that the first term in (67) is trivial and vanishes over the trace. The second term gives

$$\Omega(\boldsymbol{r}, z) = -\frac{iz^3}{4\pi^2}(f_{A,0} - f_{A,-1})\left(K_0(zr)^2 + K_1(zr)^2\right). \tag{71}$$

Using the Bessel function identities

$$I_1(x)K_0(x) + I_0(x)K_1(x) = \frac{1}{x} \tag{72a}$$

$$\boldsymbol{\nabla}\cdot(\hat{r}K_0(zr)K_1(zr)) = -z\left(K_0(zr)^2 + K_1(zr)^2\right) \tag{72b}$$

we obtain

$$f_{A,0} - f_{A,-1} = \frac{2\pi i}{z}\left(\frac{1}{zRK_0(zR)K_1(zR)}\right) \tag{73}$$

and consequently

$$\Omega(\boldsymbol{r}, z) = -\frac{1}{2\pi R}\boldsymbol{\nabla}\cdot\left(\frac{K_0(zr)K_1(zr)}{K_0(zR)K_1(zR)}\hat{r}\right). \tag{74}$$

Note that this result is identical to the one obtained in (35) and exhibits a second equivalent way to calculate $Q$ and $\rho(\boldsymbol{r})$ using (66) instead of (16) and (17).

For a vacancy of type B we only need to change $A \to B$ in (71). From (58) the only difference will be an overall sign.



\end{document}